\documentclass[a4paper,fleqn]{cas-dc}
\usepackage[numbers]{natbib}
\pdfoutput=1
\usepackage{amsmath,amsfonts}
\usepackage{array}
\usepackage[caption=false,font=normalsize,labelfont=sf,textfont=sf]{subfig}

\usepackage{textcomp}
\usepackage{stfloats}
\usepackage{url}
\usepackage{verbatim}
\usepackage{graphicx}
\usepackage{float} 
\usepackage{subfloat}
\usepackage{booktabs}  %  引入三线表宏包
\usepackage{enumerate}

\newtheorem{example}{Example}

\begin{document}
\let\WriteBookmarks\relax
\def\floatpagepagefraction{1}
\def\textpagefraction{.001}

\title [mode = title]{Vulnerable Smart Contract Function Locating Based on Multi-Relational Nested Graph Convolutional Network}

\tnotemark[1]
% Funding
\tnotetext[1]{This work is partially supported by Major science and technology projects in Anhui Province of China (202003a05020009) and National Natural Science Foundation of China (62002097).}

\author[1]{Haiyang Liu}
\ead{haiyang_liu@mail.hfut.edu.cn}

\author[1,2]{Yuqi Fan}
\cormark[1]
\cortext[1]{Corresponding author: Yuqi Fan} 
%\fnmark[1]
\ead{yuqi.fan@hfut.edu.cn}

\author[1]{Lin Feng}
\ead{fenglin@hfut.edu.cn}

\author[1]{Zhenchun Wei}
\ead{weizc@hfut.edu.cn}

\address[1]{School of Computer Science and Information Engineering, Anhui Province Key Laboratory of Industry Safety and Emergency Technology, Hefei University of Technology, Hefei 230601, China}
\address[2]{Anhui Provincial Key Laboratory of Network and Information Security, Anhui Normal Unviersity, Wuhu, 241002, China}

%\author{Haiyang Liu,
%	Yuqi Fan,
%	Lin Feng,
%	and Zhenchun Wei
%	\thanks{Manuscript received Month Day, Year; revised Month Day, Year; accepted Month Day, Year. This work is partially supported by Anhui Provincial Key R$\&$D Program of China (201904a06020024) and the open fund of Intelligent Interconnected Systems Laboratory of Anhui Province (PA2021AKSK0114), Hefei University of Technology. \textit{(Corresponding author:Yuqi Fan.)}}
%	\thanks{
%		Haiyang Liu,
%		Lin Feng,
%		Yuqi Fan,
%		and Zhenchun Wei are with the School of Computer Science and Information Engineering at Hefei University of Technology, Hefei 230601, China (e-mail: 
%		haiyang\_liu@mail.hfut.edu.cn;
%		fenglin@hfut.edu.cn;
%		yuqi.fan@hfut.edu.cn; 
%		weizc@hfut.edu.cn).}
%}

\begin{abstract}[S U M M A R Y]
The immutable and trustable characteristics of blockchain enable smart contracts to be applied in various fields. Unfortunately, smart contracts are subject to various vulnerabilities, which are frequently exploited by attackers, causing financial damage to users. Therefore, it is extremely important to perform effective vulnerability detection and locating to ensure the security of smart contracts. Deep learning has shown great advantages in smart contract vulnerability detection due to its powerful end-to-end feature learning. The previous deep learning based approaches to smart contract vulnerability detection focus on identifying whether there are vulnerabilities in a smart contract. However, this kind of detection cannot achieve fine-grained vulnerability detection, i.e., locating which function in the smart contract is vulnerable. In this paper, we study the problem of vulnerable smart contract function locating. We construct a novel Multi-Relational Nested contract Graph (MRNG) to better characterize the rich syntactic and semantic information in the smart contract code, including the relationships between data and instructions. An MRNG represents a smart contract, where each node represents a function in the smart contract and each edge describes the calling relationship between the functions. In addition, we create a Multi-Relational Function Graph (MRFG) for each function, which characterizes the corresponding function code. That is, each function is characterized as an MRFG, which corresponds to a node in the MRNG. Each MRFG uses different types of edges to represent the different control and data relationships between nodes within a function. We also propose a Multi-Relational Nested Graph Convolutional Network (MRN-GCN) to process the MRNG. MRN-GCN first extracts and aggregates features from each MRFG, using the edge-enhanced graph convolution network and self-attention mechanism. The extracted feature vector is then assigned to the corresponding node in the MRNG to obtain a new Featured Contract Graph (FCG) for the smart contract. Graph convolution is used to further extract features from the FCG. Finally, a feed forward network with a Sigmoid function is used to locate the vulnerable functions. Experimental results on the real-world smart contract datasets show that model MRN-GCN can effectively improve the accuracy, precision, recall and F1-score performance of vulnerable smart contract function locating.
\end{abstract}


\maketitle

\section{Introduction}
\label{sect:intro}
Blockchain is rising with the increasing popularity of digital currencies such as Bitcoin. The data on blockchain are recorded and maintained by all blockchain participants. The consensus mechanism makes the data on the blockchain public and transparent, such that they cannot be modified. The immutable and trustable characteristics enable blockchain to receive increasing attention \cite{RN8}.

The development of blockchain is not limited to digital cryptocurrencies and begins to expand to various fields including industry, healthcare, supply chain, etc. Blockchain has become a new distributed platform for mutual trust, on which smart contracts play an increasingly important role. A smart contract is the code stored in the blockchain system and can be executed when invoked by the users. Smart contracts can not only execute the digital transactions of blockchain, but also meet the needs of digital records, distributed computing, supply chain traceability, etc \cite{RN24, RN16, RN25}.

Smart contracts are gaining popularity. However, they are subject to various vulnerabilities, which are frequently exploited by attackers, causing huge financial losses to users. For instance, in June 2016, The DAO, a decentralized autonomous organization deployed on Ethereum, had more than \$60 million worth of digital currency stolen, and no lost digital assets can be recovered. In July 2017, Parity, a well-known multi-signature digital wallet, was maliciously attacked, resulting in a direct financial loss of \$30 million. In March 2022, the vulnerability of contract Ronin was attacked and \$615 million worth of digital assets were transferred, which was all of Ronin's funds on the blockchain. Obviously, there is an urgent need to secure smart contracts and conduct effective vulnerability detection and locating for smart contracts.
 
Some studies on smart contract vulnerability detection adopt traditional methods including symbolic execution \cite{nikolic2018finding,krupp2018teether,rodler2018sereum,torres2018osiris,RN40}, formal verification \cite{kalra2018zeus,RN53,RN7}, intermediate representation \cite{RN57,RN61}, fuzzy detection \cite{RN22,RN51}, etc. These methods can detect the vulnerabilities with established rules. However, these traditional methods rely heavily on expert knowledge and can only design limited logic rules and behavior patterns, which are prone to errors. Furthermore, the design process is also time consuming.

In recent years, deep learning has been applied to vulnerability detection in smart contracts due to its powerful end-to-end feature learning. However, most deep learning based methods analyze the smart contract code at the file level. That is, these methods classify whether there are vulnerabilities in a smart contract, by inputting the entire smart contract to the neural network for processing. This kind of methods accomplish coarse-grained vulnerability detection. In the development of smart contracts, we often need to achieve fine-grained vulnerability detection, i.e., vulnerable function locating, to facilitate the coding and debugging of smart contracts. However, the existing studies cannot locate the specific functions with vulnerabilities in a smart contract. Accordingly, these methods are inefficient when auditing and locating the functions with vulnerabilities for complex smart contracts.

To the best of our knowledge, little attention has been paid to vulnerable smart contract function locating, i.e., locating which function in the smart contract is vulnerable. A natural approach to vulnerability locating, a fine-grained vulnerability detection problem, is to use each individual function in the smart contract as input to the neural network designed for smart contract vulnerability detection. There are two concerns with this kind of methods: 

(1) The previous studies on smart contract vulnerability detection miss to exploit some of the complex relationships between instructions and data that exist in the smart contract code. These methods represent the code as a sequence or a graph. The sequence based approaches treat the code as text and characterize the execution order of statements, whereas ignoring the data flow and the control flow information. The graph based methods convert the code into a graph and extract the graph features using graph neural networks (GNNs). When converting the source code to a graph, the existing methods neglect some of the complicated relationships between nodes, such as the order of statement execution, data types of variables, control structures, etc. As a result, these methods cannot comprehensively capture the rich syntactic and semantic information in the smart contract code. When processing the graphs, most existing graph based methods use multiple stacked graph convolutional layers for feature extraction. As the layers go deeper, more indirect neighboring node features are aggregated on each node, which dilutes the features of the node itself, making it difficult to achieve a balance between local node features and global graph features.

(2) Processing each function independently ignores the fact that the invocation relationships between functions have an important impact on the vulnerability of a function. Nevertheless, a vulnerability can be caused by the inter-function relationships, as demonstrated in Example \ref{ex:FuncRelation}. 

\begin{example}
\label{ex:FuncRelation}
Figure \ref{fig:exp1} shows two smart contract functions of \textit{add} and \textit{count}. Function \textit{add} performs the addition operation on two uint16 (16-bit unsigned integer) variables, and the return variable type is also uint16. Obviously, function \textit{add} runs well independently. However, when function \textit{count} calls function \textit{add} and assigns the returned value of function \textit{add} to a unit8 variable in Line 12, an integer overflow (arithmetic vulnerability) may occur.  
\begin{figure}[h]
	\centering
	\includegraphics[width=0.48\textwidth]{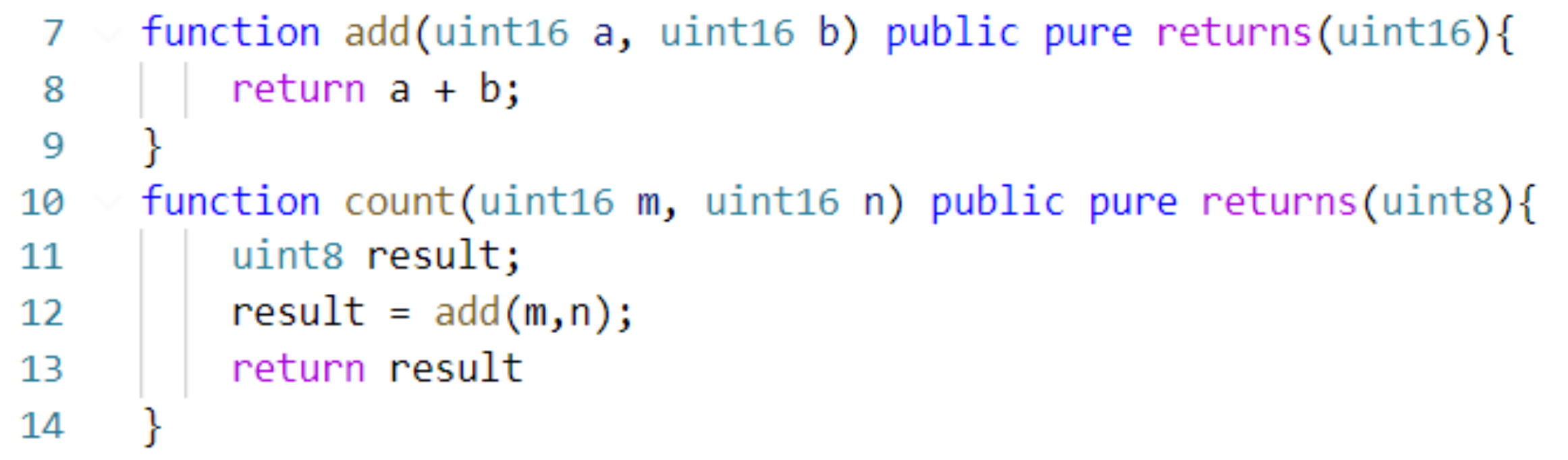}
	\caption{The impact of invocation relationships between different functions on the vulnerability of a function.}
	\label{fig:exp1}
\end{figure}

\end{example}

In this paper, we study the problem of vulnerable smart contract function locating. The contributions of this paper are as follows:

\begin{itemize}
	\item We construct a novel Multi-Relational Nested contract Graph (MRNG) to better characterize the rich syntactic and semantic information in the smart contract code, including the complex relationships between data and instructions. The nodes and edges in the MRNG represent the functions and the calling relationships between the functions in the smart contract, respectively. Each function in the smart contract code is characterized as an Multi-Relational Function Graph (MRFG), which corresponds to a node in the MRNG. Each MRFG uses different types of edges to represent the different control and data relationships between nodes within the function.
	
	\item We propose a Multi-Relational Nested Graph Convolutional Network (MRN-GCN) to learn semantic and structural features first from each MRFG and then from the MRNG. Specifically, MRN-GCN first extracts and aggregates features from each MRFG, using the edge-enhanced graph convolution network and self-attention mechanism. The extracted feature vector is then assigned to the corresponding node in the MRNG to obtain a new Featured Contract Graph (FCG) for the smart contract. Graph convolution is used to further extract features from the FCG. Finally, a feed forward network with a Sigmoid function is used to classify whether the functions are vulnerable.
	
	\item We conduct experiments on the real-world datasets containing three important and common vulnerabilities. Experimental results demonstrate that our proposed method can effectively improve the accuracy, precision, recall and F1-score of vulnerable smart contract function locating.
\end{itemize}

The rest of the paper is organized as follows. The related works are presented in Section \ref{sect:rw}. The vulnerable smart contract function locating problem is described in Section \ref{sect:problem}. MRN-GCN is proposed in Section \ref{sect:method}. We experimentally evaluate MRN-GCN in Section \ref{sect:experiment}. We discuss the limitations of our study in terms of both internal and external threats in Section \ref{sect:threats}. We finally conclude the paper in Section \ref{sect:conclusion}.

\section{Related Work}
\label{sect:rw}
The immutable and trustable characteristics of blockchain enable smart contracts to be applied in various fields such as industry, healthcare, supply chain, etc. However, smart contracts are subject to various vulnerabilities, which are frequently exploited by attackers, causing huge financial damage to users. The increasingly complex smart contracts make it difficult for current vulnerability detection methods to audit and locate the vulnerabilities. Therefore, there is an urgent need to achieve vulnerable function locating, i.e., locating which function in the smart contract is vulnerable.

To the best of our knowledge, little attention has been paid to vulnerable smart contract function locating, which is a fine-grained smart contract vulnerability detection problem. A natural approach to vulnerability locating is to use each individual function in the smart contract as input to the neural network designed for smart contract vulnerability detection. Therefore, in this section, we introduce the related works on smart contract vulnerability detection, which can be divided into two categories: traditional methods and deep learning based methods.

\subsection{Traditional Methods of Smart Contract Vulnerability Detection}
Traditional vulnerability detection methods adopt symbolic execution, formal verification, intermediate representation, fuzzy detection, etc.

\subsubsection{Symbolic Execution}
Nikoli et al. \cite{nikolic2018finding} proposed a smart contract analysis tool based on symbolic analysis, which finds security vulnerabilities through long sequences of contract invocation processes. Krupp et al. \cite{krupp2018teether} proposed a static analysis tool for smart contracts, which analyzes the bytecode to find critical execution paths to detect security issues. Rodler et al. \cite{rodler2018sereum} used dynamic taint tracking to monitor the data flow during the execution of smart contracts for the detection of reentrancy attacks. Torres et al. \cite{torres2018osiris} proposed a framework that combines symbolic execution and taint analysis to find integer errors in Ethereum smart contracts. Luu et al. \cite{RN40} proposed a static analysis tool for smart contract vulnerability detection. The tool takes bytecodes with the global state of Ethereum as input, constructs the control flow graph (CFG), and symbolically runs the smart contract to identify specific vulnerabilities through logical analysis.

\subsubsection{Formal Verification}
Kalra et al. \cite{kalra2018zeus} proposed a static analysis tool based on a formal verification approach, which uses abstract interpretation and symbolic model checking as well as constraint statements to verify the security of smart contracts. Tsankov et al. \cite{RN53} proposed Securify which analyzes the smart contract in Ethereum to show whether the smart contract is secure. For each vulnerability property, Securify defines the corresponding secure and insecure smart contract behaviors. To detect anomaly patterns, Securify symbolically analyzes the dependency graph of each smart contract, extracts the semantic attributes, and examines the key code structures, using logical conditions and scrutinization for the presence of vulnerability properties. Grishchenko et al. \cite{RN7} used a formal verification approach which transforms the bytecode of a smart contract using programming model F* and defines a series of security properties, such as invocation integrity and parameter independence, to perform the vulnerability detection.

\subsubsection{Intermediate Representation}
Feist et al. \cite{RN57} proposed Slither, a multi-vulnerability detector based on smart contract source code written in Solidity. First, Slither transforms the smart contract source code into an intermediate representation consisting of a streamlined instruction set and some semantic information. Second, Slither uses taint analysis to examine several smart contract vulnerabilities related to data dependencies. Tikhomirov et al. \cite{RN61} proposed a tool called SmartCheck which takes the source code of a smart contract as the input, converts the source code into an XML-based intermediate language, and detects the vulnerabilities by the designed behavior patterns.
 
\subsubsection{Fuzzy Detection}
Jiang et al. \cite{RN22} used fuzzy detection and runtime behavior monitoring to identify vulnerabilities during the execution of a smart contract. Liu et al. \cite{RN51} proposed an analysis tool Reguard based on fuzzing, which finds the reentrancy vulnerability in the smart contracts.
 
\subsubsection{Summary of Traditional Methods}
Traditional vulnerability detection can detect the vulnerabilities with established rules. However, these methods demand expert definition of logic rules, which takes heavy reliance on experience, professionality, and domain knowledge. Furthermore, the design of rules for logical statements is time-consuming and prone to errors. 
%Moreover, the manually-set rules in most cases are only used for specific tasks, which has significant limitations.

\subsection{Deep Learning Based Smart Contract Vulnerability Detection}
With the significant development of deep learning and its great advantages in end-to-end automatic feature learning, many researchers have adopted deep learning to obtain the internal structure information of complex data. Smart contract vulnerability detection based on deep learning can be divided into two classes: sequence based and graph based methods.

\subsubsection{Sequence Based Methods}
Tian et al. \cite{RN58} collected three kind of information including source code, annotations and account status of smart contracts, merged the above information into a sequence, and input the generated initial attributes using word embedding to a bi-directional long short memory (BiLSTM) network for feature extraction. Qian et al. \cite{RN66} treated the smart contract as a series of code fragments, mapped the information such as variables and functions within the fragments to their corresponding elements, and generated a sequence of elements for each code fragment. The sequence is then fed into a BiLSTM network for processing and classification. Xing et al. \cite{RN4} proposed a feature extraction method called slice matrix, which partitions the opcodes of a smart contract into a sequence of matrices based on the returned instructions and inputs the number of opcodes of each type in each matrix to a classifier for vulnerability detection.

The sequence based smart contract vulnerability detection methods can utilize some of the semantic information and the execution order of statements in the smart contract, but cannot represent the data flows or the control flows. Therefore, the performance of sequence based smart contract vulnerability detection methods are not satisfactory.

\subsubsection{Graph Based Methods}
Some research uses graph based methods in the vulnerability detection of codes other than smart contracts. Ma et al. \cite{RN52} extracted the abstract syntax tree (AST) from the bytecode of a smart contract. They also proposed a method to traverse and filter the semantic elements in the AST, which are compared with predefined logical rules. Xu et al. \cite{RN5} built a structural similarity matrix based on the AST as the feature vector and used a deep neural network to detect vulnerabilities in smart contracts. Cao et al. \cite{RN14} constructed the code composite graph (CCG) of the source code and proposed a bidirectional GNN for vulnerability detection. The model adds a reverse edge to each corresponding directed edge in the graph structure to form a bidirectional channel in order to make the GNN efficient. Yan et al. \cite{RN17} constructed the CFG for the code and used a deep graph convolutional neural network (DGCNN) to learn the structural and semantic information of the CFG to classify the malware. Wang et al. \cite{RN20} used the Gated Graph Neural Network (GGNN) to detect vulnerabilities in the source code. The GGNN extracts the graph features by aggregating and updating the states of nodes in the graph.

Some researchers have adopted graph based methods in smart contract vulnerability detection. Zhuang et al. \cite{RN59} constructed a normalized contract graph to represent the essential information in the smart contract source code. They used a degree-free graph convolutional neural network (DR-GCN) and a temporal message propagation (TMP) framework based on GNNs to learn the features in the CFG. Liu et al. \cite{RN19} proposed a method combining deep learning and expert knowledge. The method extracts the feature from the CFG and combines the graph feature with the designed expert patterns for smart contract vulnerability detection. Wu et al. \cite{RN47} proposed a method called Peculiar to detect the vulnerabilities in smart contracts. Peculiar generates a data flow graph from the smart contract source code and simplifies the data flow graph to extract the crucial data flow as the attribute of the smart contract. Peculiar then feeds the crucial data flow graph into CodeGraphBERT \cite{RN34} for smart contract vulnerability detection. 

The existing methods based on the graph structure have some limitations. When representing the code of each function, these methods neglect some of the complicated relationships between two nodes, such as the order of statement execution, data types of variables, control structures, etc. As a result, these methods cannot fully capture the rich syntax and semantic information in the source code. When converting the code into the graph structure, the existing methods analyze each function individually, whereas ignoring the impact of the invocation relationships between different functions on the vulnerability of a function. Furthermore, most existing methods use multiple stacked graph convolutional layers for feature extraction. As the layers go deeper, more indirect neighboring node features are aggregated on each node, which dilutes the features of the node itself, making it difficult to achieve a balance between local node features and global graph features.

\section{Problem Description}
\label{sect:problem}
In this paper, we study how to locate the vulnerable functions in a smart contract based on the smart contract source code. That is, given a smart contract, we determine whether each function in the smart contract is vulnerable. Specifically, we investigate the following three \emph{Research Questions} (RQs):
\begin{itemize}
	\item \textbf{RQ1:} How to effectively represent the syntactic and semantic information inside a smart contract function?
	\item \textbf{RQ2:} How to describe the impact of invocation relationships between functions in a smart contract on the vulnerability of functions?
	\item \textbf{RQ3:} How to design a neural network that can effectively extract the syntactic and semantic features within and between smart contract functions?
\end{itemize}

We investigate three typical vulnerabilities, i.e., Arithmetic, Reentrancy and Timestamp Dependency, which are common in smart contracts and can cause significant financial damage to users \cite{RN53,RN22,RN59,RN47}. For instance, in the BeautyChain (BEC) Attack occurred in April 2018, the attackers exploited the integer overflow behavior (Arithmetic vulnerability) to steal a large amount of BEC tokens. This attack led to a massive sell-off of BEC tokens in the market, causing a devastating blow to BEC market trading. In the famous DAO attack, Reentrancy was exploited by attackers. Timestamp Dependency can be utilized by malicious miners for their own advantage, allowing them to always win the benefits.

\paragraph{Arithmetic} An arithmetic vulnerability exists when we assign to the variable a value beyond the valid value range. Arithmetic vulnerabilities are of two types, i.e., overflow where the variable exceeds the maximum permissible value, and underflow where the variable is less than the minimum permissible value. The arithmetic vulnerability can cause smart contracts to deliver wrong outputs and thus is a severe threat to the security.

\begin{figure}[h]
	\centering
	\includegraphics[width=0.48\textwidth]{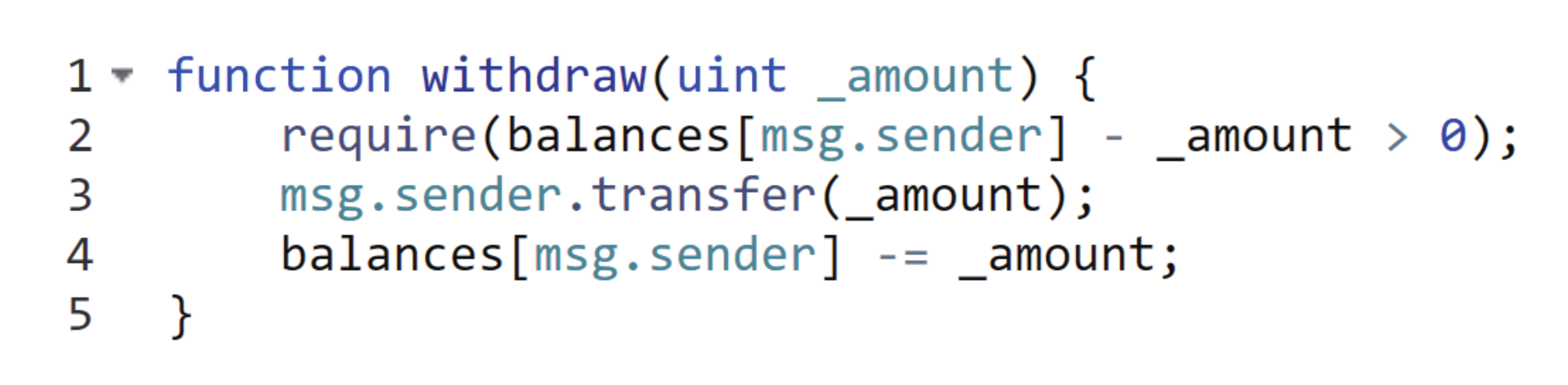}
	\caption{Example of a smart contract function with the arithmetic vulnerability.}
	\label{fig:arithmetic_example}
\end{figure}

Figure \ref{fig:arithmetic_example} shows a real smart contract function with the arithmetic vulnerability. The function first calculates whether the result of the subtraction operation is greater than zero. Note that the minimum permissible value of an unsigned integer is 0. When \_amount is a value greater than the balance, underflow occurs, which enables the attacker to withdraw almost all of the digital assets.

\paragraph{Reentrancy} Reentrancy is a vulnerability of great concern existing in Ethereum smart contracts, which was exploited by attackers in The DAO resulting a total loss beyond 60 million dollars. An Ethereum smart contract can call another smart contract during execution, and the one being called must wait until the call is finished. There is a special mechanism in smart contracts, called fallback, which has no function name and takes no arguments. The fallback function will be invoked in two scenarios when the function call does not match any functions in the called smart contract or the ether (the dedicated cryptocurrency used in Ethereum) is received by the caller. The recipient of the call can exploit this intermediate status to steal digital currency. 

\begin{figure}[h]
	\centering
	\includegraphics[width=0.48\textwidth]{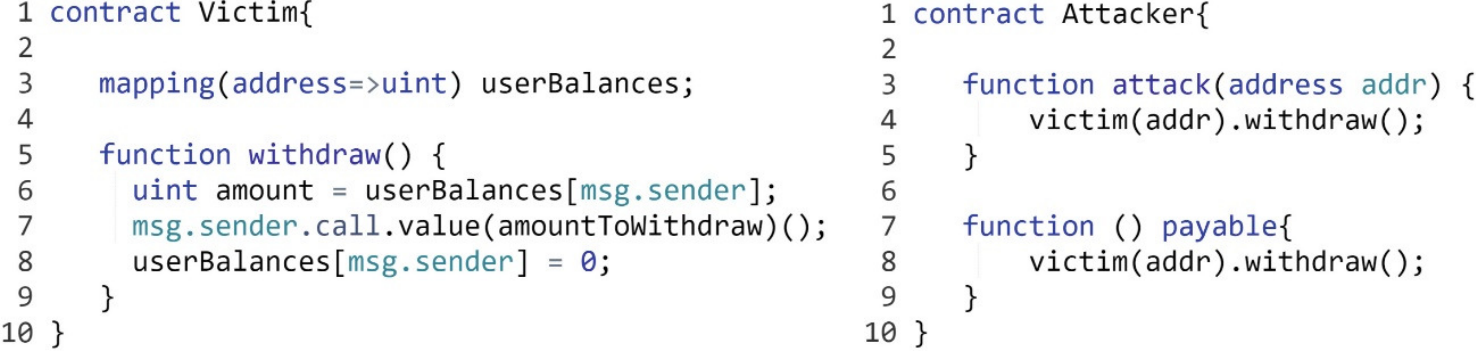}
	\caption{Example of a smart contract with the reentrancy vulnerability.}
	\label{fig:reentrancy_example}
\end{figure}

Figure \ref{fig:reentrancy_example} shows a smart contract function with the reentrancy vulnerability. There are two smart contracts of Attacker and Victim in Figure \ref{fig:reentrancy_example}. Attacker attempts to steal the ether from contract Victim, a simplified version of digital wallet with a reentrancy vulnerability, by exploiting the fallback mechanism. To be specific, Attacker executes its function attack() to invoke function withdraw() in Victim. Victim will then transfer a certain amount of ether to Attacker. Once the ether is received, the fallback function in Attacker will be executed. As we can see, the balance of Attacker’s account has not yet been set to 0 by Victim at that time, and hence Attacker can repeatedly invoke function withdraw() until the ether held by Victim drops to 0.

\paragraph{Timestamp dependency} Timestamp dependency is another well-known Ethereum smart contract vulnerability which is related to the timestamp in the blockchain system. When a smart contract takes the block timestamp as a dependency condition to trigger certain critical operations like ether transfer, some malicious miners may manipulate it to modify the timestamp for illegal benefits.

\begin{figure}[h]
	\centering
	\includegraphics[width=0.48\textwidth]{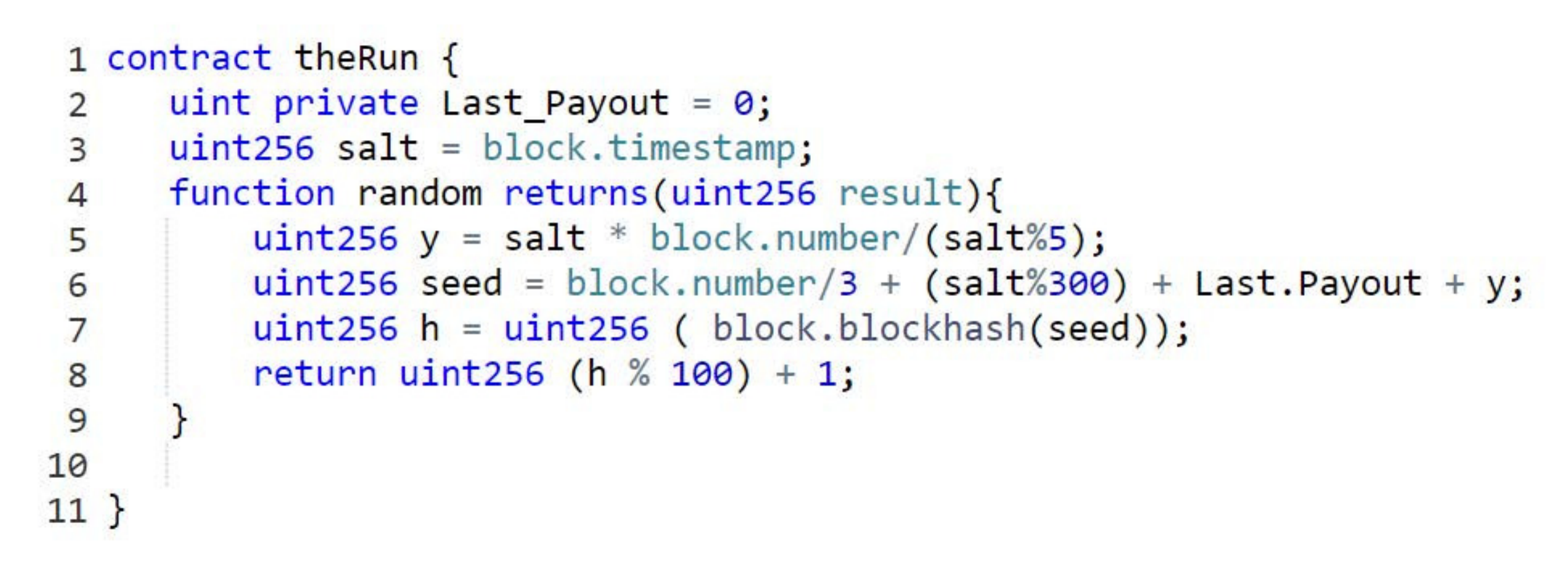}
	\caption{Example of a smart contract with the timestamp dependency vulnerability.}
	\label{fig:timestampDependency_example}
\end{figure}

Figure \ref{fig:timestampDependency_example} shows a smart contract function with the timestamp dependency vulnerability. Smart contract theRun uses a set of rules based on the current block timestamp to choose who will win the bonus. In theRun, the variable \textit{h} defined in Line 7 is the hash value of a certain block in the blockchain, which is used to decide the winner. The selection of a block is determined by the variable \textit{seed} defined in Line 6. Three variables decide the seed value, including \textit{block number}, \textit{last payout}, and \textit{timestamp}. Among them, block number and last payout are determined values recorded on blockchain, while timestamp is decided by the miner. Normally, the timestamp is set as the current time of the miner’s local system. However, the miner can vary the timestamp value by roughly 900 seconds following the consensus protocol. Therefore, by choosing the timestamp, the miner can calculate the result in advance and manipulate the outcome of timestamp-dependent smart contracts to obtain the final bonus.

\section{Proposed Method}
\label{sect:method}
In this section, we answer the research questions and present the framework and details of the proposed method.

\textbf{RQ1: How to effectively represent the syntactic and semantic information inside a smart contract function?} 

\emph{Answer:} We construct a new Multi-Relational Function Graph (MRFG) to better represent the semantic and structural information in a smart contract function. 

\textbf{RQ2: How to describe the impact of invocation relationships between functions in a smart contract on the vulnerability of functions?} 

\emph{Answer:} We construct a novel Multi-Relational Nested contract Graph (MRNG) to characterize the invocation relationships between functions in a smart contract.

\textbf{RQ3: How to design a neural network that can effectively extract the syntactic and semantic features within and between smart contract functions?}

\emph{Answer:} We propose a vulnerable function locating model, i.e., Multi-Relational Nested Graph Convolutional Neural Network (MRN-GCN), to learn semantic and structural features from the MRNG.

\begin{figure*}[hbtp]
	\centering
	\includegraphics[width=6.2in]{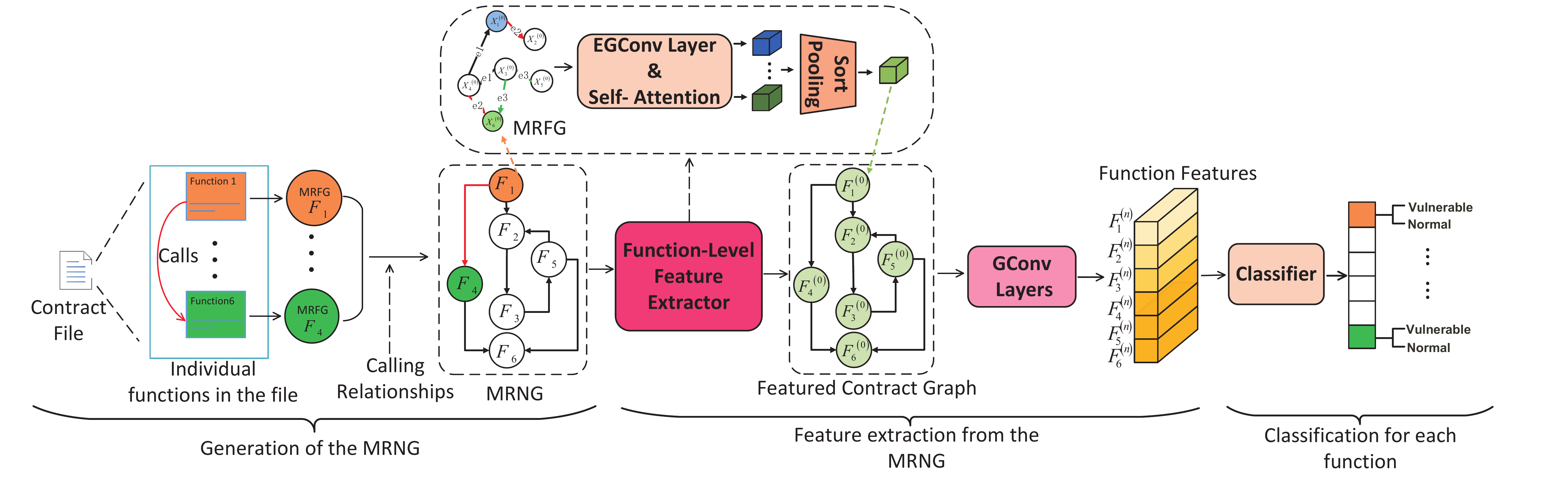}
	\caption{The framework of our proposed method. }
	\label{fig:framework}
\end{figure*}

\textbf{\emph{Framework.}} As shown in Fig. \ref{fig:framework}, our method consists of three stages.

\begin{enumerate}[\textbf{Stages} 1.] 
	\item Generate an MRNG for each smart contract, where each node represents a function and each edge describes the invocation relationship between two corresponding functions. In addition, each function is characterized as an MRFG, which corresponds to a node in the MRNG.
	
	\item Propose model MRN-GCN to learn semantic and structural features first from each MRFG and then from the MRNG.
	
	\item Use a classifier to classify each node in the MRNG to determine the validity of the corresponding function.
\end{enumerate}

\subsection{Multi-Relational Nested Contract Graph}
A commonly used way of representing the smart contract code is the abstract syntax tree (AST), which can provide the syntax and semantic information of the smart contract source code. However, the AST contains some redundant information unrelated to vulnerabilities, while lacking the representation of function structures, the order of statement execution, control flows and data flows. In this section, we first propose a Multi-Relational Function Graph (MRFG) to characterize the code of each smart contract function by improving the AST. We then construct the MRNG for the smart contract based on the invocation relationships between the functions.

The construction of MRNG includes three steps: building the AST of each function in a smart contract, generating the MRFG based on the AST, and creating the MRNG.

\subsubsection{Generation of the AST}
We build the AST of a smart contract function using python-solidity-parser \cite{RN49} which is a state-of-the-art parsing tool for \emph{solidity}, a language used in developing smart contracts.

\begin{figure}[h]
	\centering
	\includegraphics[width=0.48\textwidth]{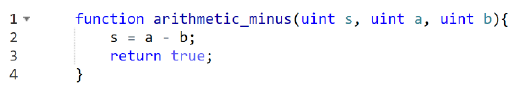}
	\caption{Example of a smart contract.}
	\label{fig:sc_exp}
\end{figure}

\begin{figure*}[hbtp]
	\centering
	\includegraphics[width=6.2in]{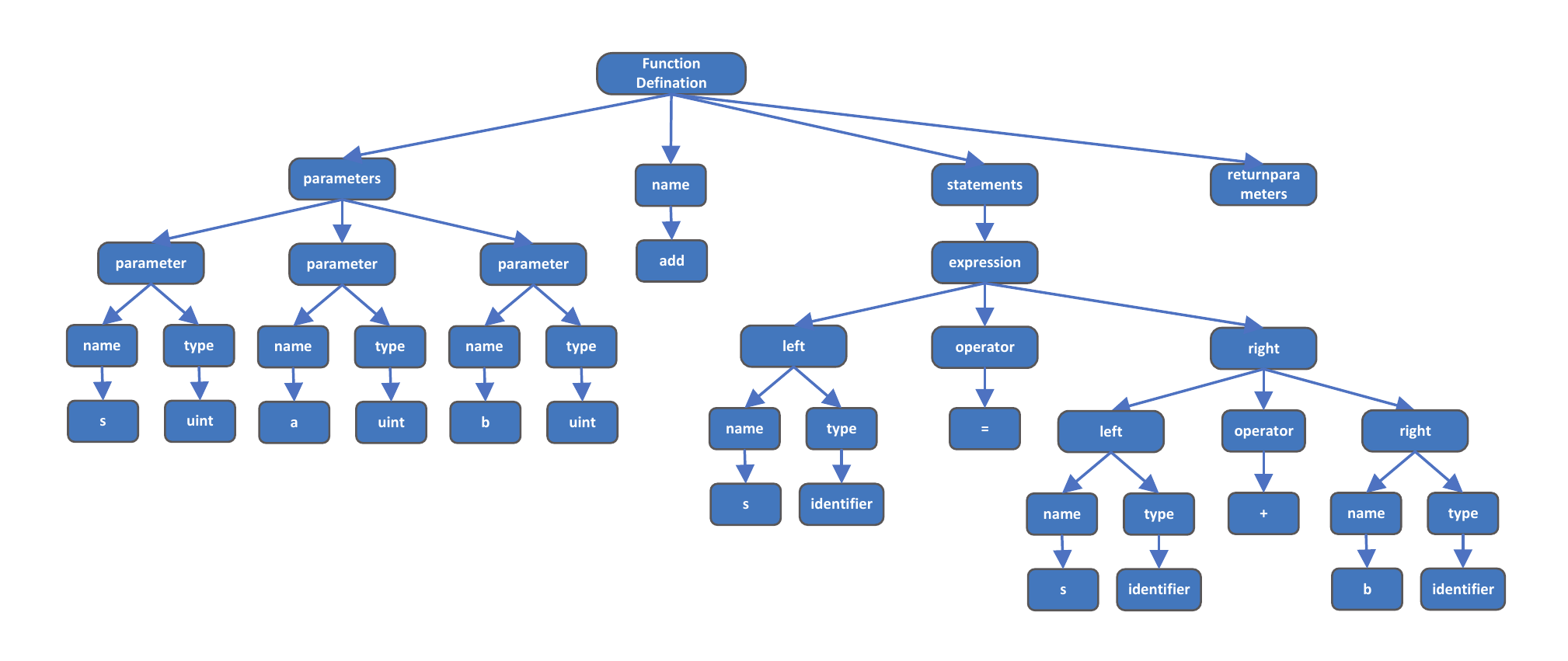}
	\caption{The AST of the given smart contract function shown in Figure \ref{fig:sc_exp}.}
	\label{fig:AST_exp}
\end{figure*}

Figure \ref{fig:sc_exp} shows a smart contract, which performs a minus operation on two unsigned integer (uint) variables \emph{a} and \emph{b} and assigns the output to variable \emph{s}. The generated AST corresponding to this smart contract function is shown in Fig. \ref{fig:AST_exp}.
%The function cannot handle the output that is less than the minimum permissible value and resulting in an integer underflow, since \emph{s} is also an unsigned integer. 

\subsubsection{Generation of the MRFG}
To address the problem that the AST lacks the data and the control flow information, Cao et al. \cite{RN14} converted an AST into a composite code graph (CCG) by adding edges between the nodes. However, during the training process, the network cannot differentiate how the nodes are connected, since the data and the control flows are represented by the same edges. Therefore, we define different kinds of edges to describe the different relationships between nodes in order to better express the semantic and structural information in smart contract functions. The edges can be categorized into the following types shown in Table \ref{tab:edgeTypes}.

\begin{table}[htbp]
	\centering  
	\caption{Types and examples of edges in the multi-relational function graph.} 
	\label{tab:edgeTypes}  % 用于索引表格的标签
	\resizebox{\linewidth}{!}{
	\begin{tabular}{c|c}
		\hline
		Types & Examples \\ \hline
		Data   Type & uint, uint8, uint64, boolean, etc. \\
		Control   Info & sequential,   if, while, etc. \\
		Fields & left, right,   operation, function call, etc. \\
		Data   Flow & compute   from, value from \\
		Fallback & fallback \\ \hline
	\end{tabular}	
	}	
\end{table}

We first perform a depth-first traversal of the AST to drop the elements that are irrelevant to vulnerabilities, including the field of attributes (name, type, body, parameter etc.), the field of flags (isStateVar, isIndexed, storageLocation, visibility, stateMutability etc.), and the field of description (TypeName, ElementaryTypeName, ExpressionStatement, etc.). We then add edges between the remaining nodes.

\begin{figure*}[h]
	\centering
	\includegraphics[width=6.0in]{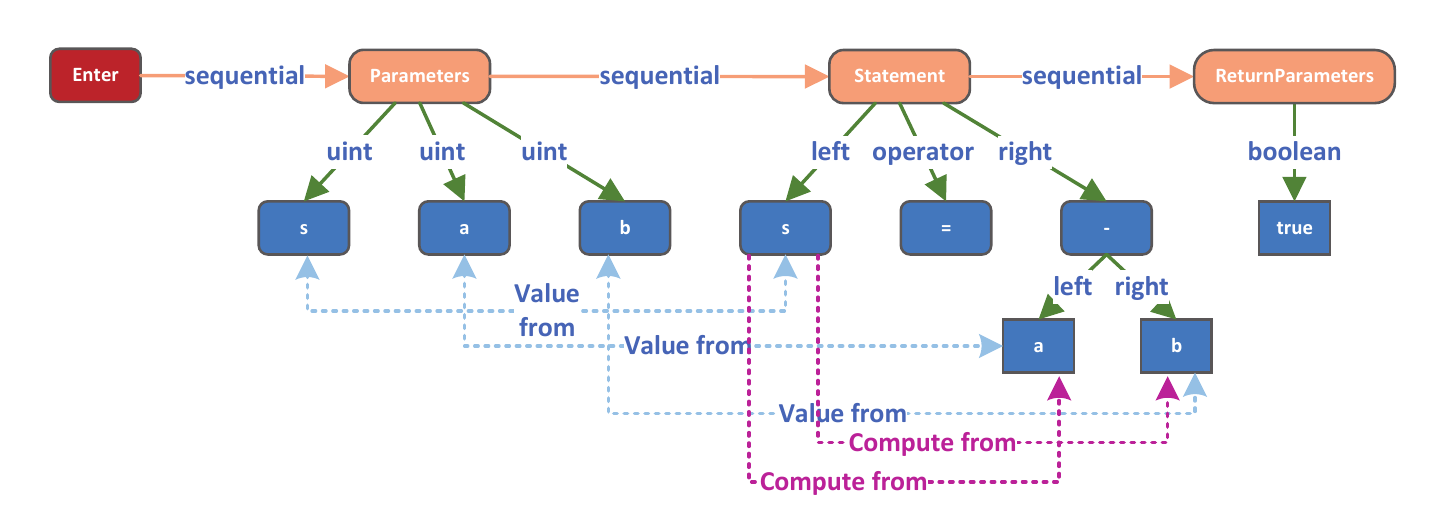}
	\caption{The MRFG of the given AST shown in Figure \ref{fig:AST_exp}.}
	\label{fig:MRFG_exp}
\end{figure*}

The code element organization of a smart contract function is as follows: the entry to the function, the list of input parameters, one or more statements, and the list of return parameters. The nodes corresponding to the elements are connected with \emph{sequential} edges to express the execution order of the statements in the smart contract function. In the subtrees of \emph{Parameters} and \emph{returnParameters} nodes, we add the edges of the corresponding data type between the \emph{Parameters}  node and each variable node. Under the \emph{expression} node, the control flow and structure information such as \textit{left, right, if} and \textit{operation} represented by the nodes in the AST are converted into the form of edges added between the corresponding nodes. In order to express the data flow in the smart contract function, we add \textit{compute from} edges and \textit{value from} edges. When the result of an expression is assigned to a variable, \textit{compute from} edges are added from this variable to all nodes that participate in the expression. When a variable is referenced by an identifier, a \textit{value from} edge is added from the identifier node to the variable node in the subtree of \textit{Parameters}. To explicitly represent the fallback mechanism in smart contracts, we add a \textit{fallback} edge which points to node \textit{entry} from the function related to the transfer of digital currency. The MRFG generated from the AST in Figure \ref{fig:AST_exp} is shown in Figure \ref{fig:MRFG_exp}.

\begin{figure*}[h]
	\centering
	\includegraphics[width=6.0in]{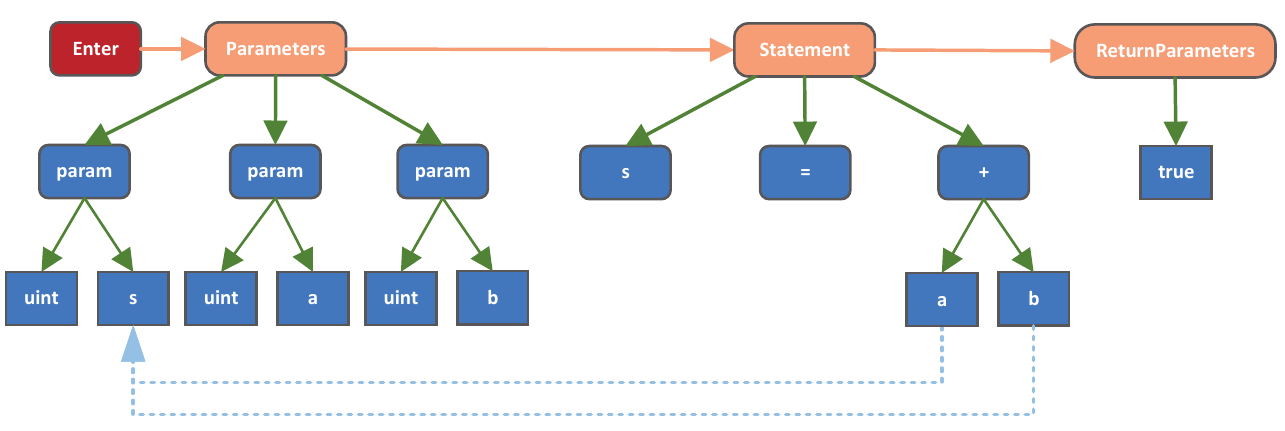}
	\caption{The CCG of the given smart contract function shown in Figure \ref{fig:sc_exp}.}
	\label{fig:CCG_exp}
\end{figure*}

\begin{figure*}[h]
	\centering
	\includegraphics[width=6.0in]{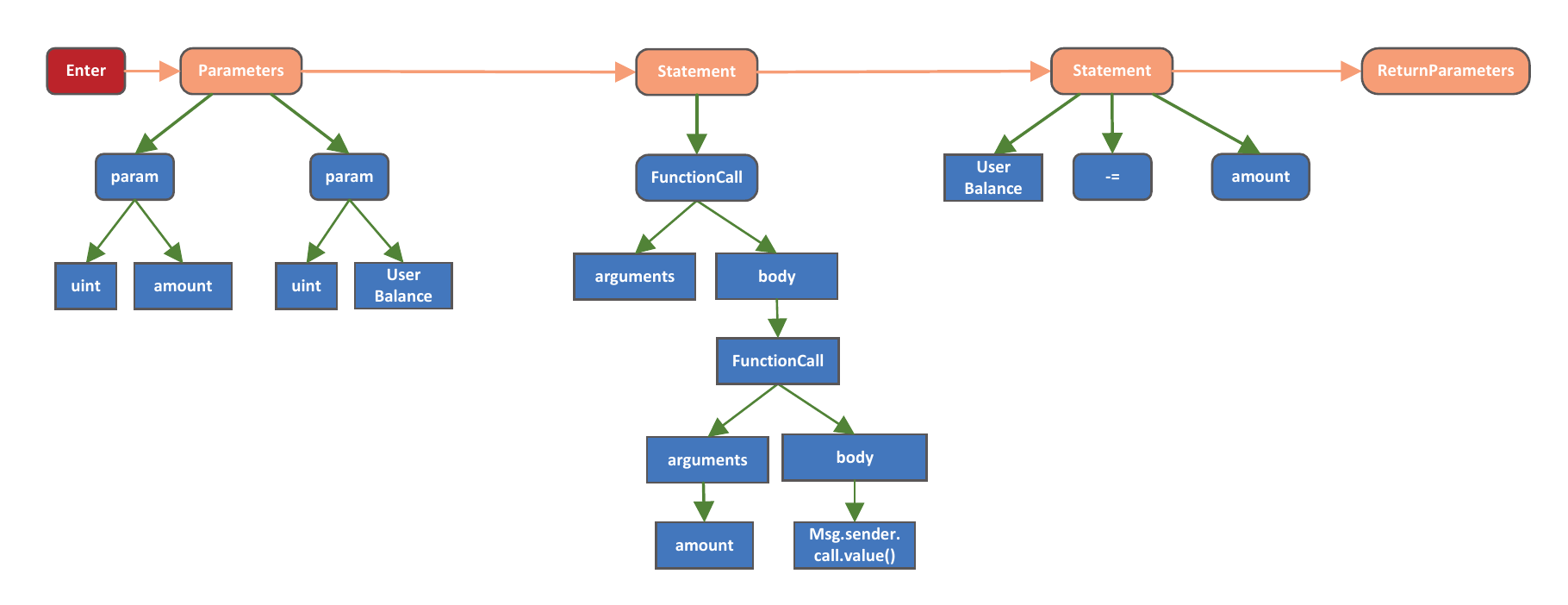}
	\caption{The CCG of the given smart contract function shown in Figure \ref{fig:reentrancy_exp}.}
	\label{fig:reentrancyCCG_exp}
\end{figure*}

\begin{figure*}[hbt]
	\centering
	\includegraphics[width=6.0in]{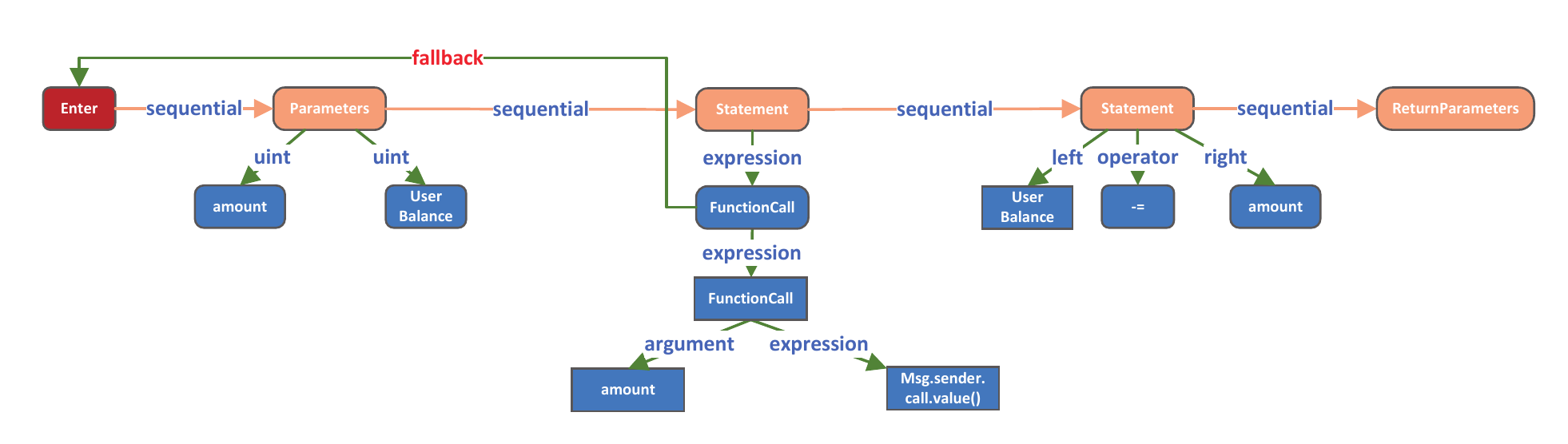}
	\caption{The MRFG of the given contract smart function shown in Figure \ref{fig:reentrancy_exp}.}
	\label{fig:reentrancyMRFG_exp}
\end{figure*}

We further present two examples to explain the advantages of the MRFG. Figures \ref{fig:MRFG_exp} and \ref{fig:CCG_exp} respectively illustrate the MRFG and the CCG corresponding to the smart contract function shown in Fig. \ref{fig:sc_exp}. 

\begin{example}
In the CCG shown in Fig. \ref{fig:CCG_exp}, the two child nodes \textit{uint} and \textit{s} of node \textit{param} are considered as equal nodes. However, node \textit{uint} should be treated as a property of variable \textit{s}. Using an edge to express the data type in the MRFG can better describe the variable declaration in the function. Besides, node \textit{a} and node \textit{b} are connected to the parent node with no difference, when statement \textit{a-b} is described in CCG. In contrast, the \textit{left} and the \textit{right} edges in the MRFG allow different representations of statements of \textit{a-b} and \textit{b-a}, enabling the features of node \textit{a} and node \textit{b} to be propagated in a semantic way. In terms of data flows, the MRFG also provides a more detailed description of how data are delivered between the variables using \textit{compute from} edges and \textit{value from} edges.
\end{example}

\begin{figure}[hbt]
	\centering
	\includegraphics[width=0.48\textwidth]{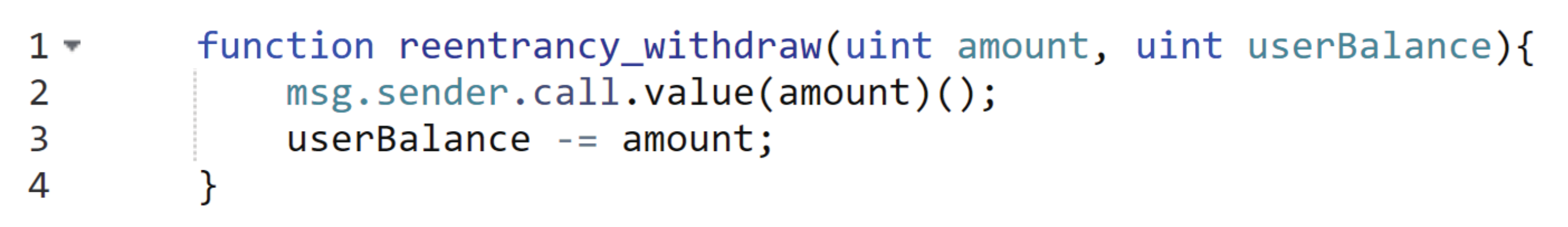}
	\caption{Example of a smart contract with the reentrancy vulnerability.}
	\label{fig:reentrancy_exp}
\end{figure}

\begin{example}
Figures \ref{fig:reentrancyCCG_exp} and \ref{fig:reentrancyMRFG_exp} show the CCG and the MRFG generated for a function with the reentrancy vulnerability, respectively. The function is shown in Fig. \ref{fig:reentrancy_exp}. In the CCG, the edges can only indicate whether two nodes are related to each other but cannot tell the execution order. On the contrary, the MRFG defines the additional edge \textit{sequential} and uses positional encoding to express the ordering information of statements. In addition, the CCG lacks the description of the fallback mechanism, while the MRFG uses the \textit{fallback} edge to represent the specific fallback mechanism in smart contracts, allowing the neural network to better understand the reentrancy vulnerability.
\end{example}

%After the MRFG is generated, the different types of nodes and edges in the MRFG are mapped to different integers as the features, which are fed into the neural network.

\subsubsection{Generation of the MRNG}
For each smart contract, we further construct the nested contract graph by capturing the function call relationships in the smart contract. 

In the MRNG, each node represents a function in the smart contract, and each edge characterizes the invocation relationship between two functions. Each node contains an MRFG corresponding to the function, in which the nodes and the edges represent the semantic elements and the relationship between two elements, respectively.

\subsection{Multi-Relational Nested Graph Convolutional Network}
We propose a novel model MRN-GCN to identify whether the functions are vulnerable by analyzing each node in the MRNG. We map each kind of nodes and edges in the MRFG to a unique integer, which is used as the MRFG feature. Model MRN-GCN first extracts and aggregates the feature of each MRFG using an edge-enhanced graph convolutional network \cite{RN30} and self-attention to obtain the feature vector, which is assigned to the corresponding node in MRNG. Accordingly, we obtain a new featured contract graph (FCG). Model MRN-GCN then extracts the features among the functions from the FCG with a GNN. Finally, model MRN-GCN uses a feed forward network to locate the vulnerable functions in the smart contract. 

\begin{figure*}[htbp]
	\centering
	\captionsetup{justification=centering}
	\includegraphics[width=7in]{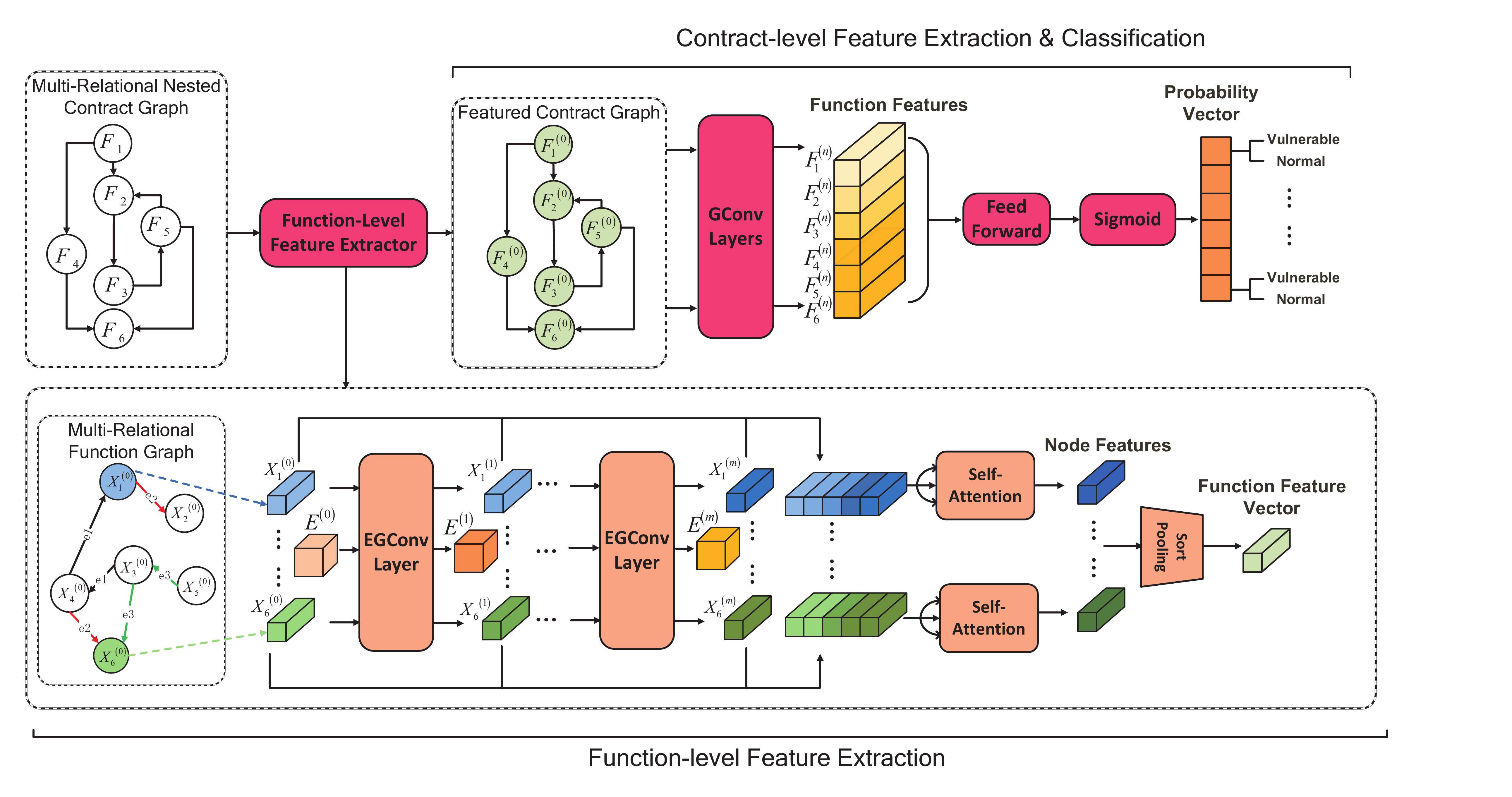}
	\caption{Diagram of multi-relational nested graph neural network.}
	\label{fig:MRN-GCN}
\end{figure*}

The MRN-GCN model shown in Fig. \ref{fig:MRN-GCN} consists of four parts: edge-enhanced graph convolutional module, self-attention module, nested graph convolutional module and classification module.

\subsubsection{Edge-Enhanced Graph Convolutional Module}
In this module, we use the edge-enhanced graph convolutional network to extract features from the generated MRFGs. We first apply word embedding to the nodes and edges, respectively, such that the semantic elements and the different types of edges mapped as integers are transformed into fixed-sized real value vectors. The parameters of word embedding are randomly initialized and then updated during the training process. We use positional encoding to the sequential-type edges to make better use of the ordering information of the statements in a smart contract function. The positional encoding is added to the embedding vector and defined as:

\begin{equation}
	\left\{ \begin{array}{*{35}{l}}
		PE_{i,2p}=\sin (i/{{10000}^{2p/{{d}_{\text{embedding}}}}})  \\
		PE_{i,2p+1}=\cos (i/{{10000}^{2p/{{d}_{\text{embedding}}}}})  \\
	\end{array} \right.
\end{equation}
where $i$ denotes the $i$-th position in an opcode sequence. $2p$ and $2p+1$ represent the $2p$-th and the $(2p\!+\!1)$-th dimensions in the embedding vector of an opcode, respectively. $d_{\text{embedding}}$ is the dimension of the embedding vector of an opcode. $PE_{i,2p}$ and $PE_{i,2p+1}$ are the encoding values of the $2p$-th and the $(2p\!+\!1)$-th dimensions in the embedding vector of the $i$-th position in an opcode sequence, respectively. That is, every dimension in the positional encoding matches a sinusoidal signal which wavelength grows geometrically from 2$\pi$ to 20,000$\pi$.

For an MRFG with $N$ nodes, we use the feature vectors obtained from word embedding and positional encoding to obtain the feature matrices of the nodes and the edges, denoted as ${{X}^{0}}\in {{\mathbf{R}}^{N\times F_{0}}}$ and $E^{0}\in {{\mathbf{R}}^{N\times N \times P}}$, respectively, where $F_{0}$ and $P$ are the dimensions of the feature vectors of the nodes and the edges, respectively. 

The propagation of the edge-enhanced graph convolutional layer is defined as follows:
\begin{equation}
\label{eq:Xl}
	X^{l}=\sigma \left[ \underset{p=1}{\overset{P}{\mathop{\|}}}\,\left( f_{\cdot \cdot p}^{l}\left( X_{{}}^{^{l-1}},E_{\cdot \cdot p}^{l-1} \right)X_{{}}^{^{l-1}}W_{{}}^{^{l}} \right) \right]
\end{equation}
where $X^{l} \in {{\mathbf{R}}^{N\times {{F}_{hidden}}}}$ and ${{X}^{l-1}}\in {{\mathbf{R}}^{N\times {{F}_{hidden}}}}$ are the outputs of the $l$-th and the $(l\!-\!1)$-th edge-enhanced graph convolutional layers, respectively. $\sigma$ denotes the activation function, $\|$ represents the concatenation operation, and $f$ is a function which output is an attention coefficient matrix with dimension $N\times N\times P$. $W^{l}$ denotes the parameter mapping from the $l$-th layer to the $(l\!-\!1)$-th layer.

The attention coefficient $f_{ijp}^{l}$ between node $i$ and node $j$ is calculated as
\begin{equation}
	f_{ijp}^{l}=\exp \left[ \sigma \left( X_{i\cdot }^{^{l-1}}{{W}^{l}}\|X_{j\cdot }^{^{l-1}}{{W}^{l}} \right) \right]E_{ijp}^{l-1}
\end{equation}
where ${{W}^{l}}$ is the same parameter mapping as that in Eq. \eqref{eq:Xl}.

The attention coefficient matrix is used as the new edge feature to participate in the next layer of the network as follows:
\begin{equation}
	E^{l}=f^{l}
\end{equation}

\subsubsection{Self-Attention Module}
As the layers of the graph neural network go deeper, each node will aggregate more indirect neighboring node features on the node, which dilutes the features of the node itself and makes it difficult to achieve a balance between the local node features and global graph features. As a result, the obtained features cannot be used well for vulnerable function locating.

In this module, we concatenate the feature matrix obtained from each edge-enhanced graph convolutional layer and perform self-attention to extract and aggregate the features from different layers. The self-attention is calculated as follows:
\begin{equation}
	\text{Attention}(Q,K,V)=\text{softmax}(\frac{Q{{K}^{T}}}{\sqrt{{{d}_{k}}}})V
\end{equation}
where $Q$, $K$ and $V\in {{\mathbf{R}}^{length\times {{d}_{k}}}}$ denote query, key and value vectors, respectively. ${{d}_{k}}$ is the dimension of the query, key and value vectors.

We then apply graph sort pooling to sort the nodes in the non-ascending order of the features and keep the top $K'$ nodes to obtain the most effective features.

\subsubsection{Nested Graph Convolutional Module}
In this module, we assign the feature vector of each function to the corresponding node in the MRNG to obtain a new graph FCG. We then use the graph convolutional network to further extract features among the functions from the FCG.

For the FCG with ${N}_{F}$ nodes, we denote the graph feature matrix as $Z^{0} \in {{\mathbf{R}}^{{{N}_{F}} \times c_{0}}}$, where $c_{0}$ is the dimension of the node features. We use multiple staked graph convolutional layers as follows:
\begin{equation}
	{{Z}^{t+1}}=f({{\tilde{D}}^{-1}}\tilde{A}{{Z}^{t}}{{W}^{t}})
\end{equation}
where $f$ is the activation function. $\tilde{A}$ is the adjacency matrix with self-loop, $\tilde{D}$ is the corresponding degree matrix, and $Z^{t} \in {{\mathbf{R}}^{N\times c_{t}}}$ is the output of the $t$-th graph convolutional layer. $c_{t}$ is the number of output channels of the $t$-th graph convolutional layer, and $W^{t} \in {{\mathbf{R}}^{{c}_{t} \times c_{t+1}}}$ denotes the parameter mapping from $c_{t}$ channels to $c_{t+1}$ channels.

\subsubsection{Classification Module}
After extracting the features with the previous modules, we use the convolutional layer, the pooling layer, and the fully connected layer to obtain the final feature vector in which each dimension represents a function in the smart contract. We finally use a Sigmoid layer to identify whether each function is vulnerable.

\section{Experiment}
\label{sect:experiment}

\subsection{Experimental Setup}
%\subsubsection{Dataset and Performance Metrics }
We use the real-world smart contracts from SmartBugs \cite{RN27} as the datasets. Our datasets include three different vulnerabilities: arithmetic, reentrancy and timestamp dependency. Specifically, as shown in Table \ref{tab:dataset_func}, there are 12505 smart contract files with 15635 vulnerable functions and 16553 normal functions in the arithmetic vulnerability dataset, 3116 smart contract files with 5722 vulnerable functions and 6702 normal functions in the reentrancy vulnerability dataset, and 587 smart contract files with 625 vulnerable functions and 803 normal functions in the timestamp dependency vulnerability dataset. We randomly select 80\% of the dataset as the training set, 10\% of the dataset as the validation set and the remaining 10\% of the dataset as the test set. For each method, 30 rounds of experiments are run and the average of the 30 results is taken as the final result. 

%\begin{table}[htbp]
%	\centering  % 显示位置为中间
%	\caption{The division of the Arithmetic,Reentrancy,Timestamp vulnerability datasets.}  % 表格标题
%	\label{dataset_split}  % 用于索引表格的标签
%	\resizebox{\linewidth}{!}{
%	\begin{tabular}{ccccc}
%		\toprule
%		DataSet & Arithmetic & Reentrancy & Timestamp  \\
%		\midrule
%		Training set (80\%)  & 10004 & 2493 & 470 \\
%		Validation set (10\%) & 1251 & 312 & 59 \\
%		Test set (10\%)  & 1250 & 311 & 58 \\
%		Total	& 12505 & 3116 & 587 \\
%		\bottomrule
%	\end{tabular}
%	}
%\end{table}

\begin{table}[!h]
	\centering  % 显示位置为中间
	\caption{The function information contained in the arithmetic, reentrancy, timestamp dependency vulnerability datasets.}  % 表格标题
	\label{tab:dataset_func}  % 用于索引表格的标签
	\resizebox{\linewidth}{!}{
		\begin{tabular}{ccccc}
			\toprule
			DataSet & Arithmetic  & Reentrancy & Timestamp Dependency \\
			\midrule
			Normal functions & 16553 & 6702 & 803 	\\
			Vulnerable functions & 15635 & 5722 & 625	\\
			Contract files & 12505 & 3116 & 587 	\\
			\bottomrule
		\end{tabular}
	}
\end{table}

We evaluate our proposed method using 4 commonly used performance metrics in classification tasks: accuracy, recall, precision and F1-score as calculated by Eqs. \eqref{eq:accuracy}$\sim$\eqref{eq:F1}, where \textit{TP}, \textit{FP}, \textit{FN} and \textit{TN} represent the numbers of true positive samples, false positive samples, false negative samples and true negative samples, respectively. 

\begin{equation}
	\label{eq:accuracy}
	accuracy=\frac{TP+TN}{TP+FP+TN+FN}
\end{equation}

\begin{equation}
	\label{eq:precision}
	precision=\frac{TP}{TP+FP}
\end{equation}

\begin{equation}
	\label{eq:recall}
	recall=\frac{TP}{TP+FN}
\end{equation}

\begin{equation}
	\label{eq:F1}
	F1\!-\!score=\frac{2\times precision\times recall}{precision+recall}
\end{equation}

%\subsubsection{Implementation Details}
All the experiments are run on a desktop with an Intel i7-12700KF 3.6GHz CPU, an NVIDIA GeForce 3080 GPU and 32GB memory. The MRN-GCN model is written in Python and run under PyTorch.

We use grid search to find the best parameters for MRN-GCN, and the selected hyperparameters of MRN-GCN are shown in Table \ref{tab:model_parms}. The code we work on is at https://github.com/coder644/MRN-GCN.
\begin{table}[!h]
	\centering  % 显示位置为中间
	\caption{Settings of hyperparameters of MRN-GCN.}  % 表格标题
	\label{tab:model_parms}  % 用于索引表格的标签
	%\resizebox{\linewidth}{!}{
		\begin{tabular}{p{5cm}|p{2.5cm}}
			\toprule
			Parameter & Value  \\
			\midrule
			Epoch  & 100  \\
			Batch size  & 32 \\
			Learning rate & 0.002 \\
			Optimizer	& SGD  \\
			Momentum	& 0.0005  \\
			Loss function	& Cross entropy  \\
			EGCN layers		& 16  \\
			Dropout		& 0.2  \\
			Number of heads in self-attention  & 4  \\
			\bottomrule
		\end{tabular}
	\end{table}
%\begin{table}[htbp]
%	\centering  % 显示位置为中间
%	\caption{Hyperparameter settings in experiments.}  % 表格标题
%	\label{tab:model_parms}  % 用于索引表格的标签
%	\resizebox{\linewidth}{!}{
%		\begin{tabular}{c|c|c|c}
%			\toprule
%			DataSet & Arithmetic & Reentrancy & Timestamp  \\
%			\midrule
%			Epoch  & 100 & 100 & 100 \\
%			Batchsize  & 32 & 32 & 32 \\
%			Learning rate & 0.002 & 0.005 & 0.005 \\
%			Optimizer	& SGD & SGD & SGD \\
%			Momentum	& 0.0005 & 0.0005 & 0.0005 \\
%			Loss function	& CrossEntropy & CrossEntropy & CrossEntropy \\
%			EGCN layers		& 16 & 16 & 16 \\
%			Dropout		& 0.2 & 0.2 & 0.2 \\
%			Self-attention heads & 4 & 4 & 4 \\
%			\bottomrule
%		\end{tabular}
%	}
%\end{table}

\subsection{Experimental Results}
\subsubsection{Performance Comparison of Different Vulnerable Smart Contract Function Locating Methods}
We compare the proposed model MRN-GCN with the following four popular or state-of-the-art methods: 

(1) Transformer \cite{vaswani2017attention}: A classical and popular method for processing sequence data. Transformer mines the long-term dependencies within sequences by self-attention. In our experiments, we use the Transformer encoder followed by the same classifier in our method to conduct the classification of smart contract functions.

(2) DGCNN \cite{RN17}: A classical and popular method for malicious code detection using graph convolutional networks. DGCNN improves the performance of graph classification tasks by establishing the connection between GNN and CNN through the SortPooling layer. We run DGCNN on the MRNGs constructed via the proposed method in this paper. 

(3) Peculiar \cite{RN47}: One of the state-of-the-art algorithms for smart contract vulnerability detection. Peculiar builds a feature graph based on the crucial data flows in the smart contract and feeds the graph into CodeGraphBERT \cite{RN34} for smart contract classification.

(4) TMP \cite{RN59}: One of the state-of-the-art algorithms for smart contract vulnerability detection. TMP constructs a normalized contract graph based on the data and control dependencies in the smart contract source code. TMP then uses a temporal graph message propagation framework to learn the features in the normalized contract graph. 

The experimental results of arithmetic, reentrancy and timestamp dependency vulnerability locating tasks are shown in Tables \ref{tab:comparisonAri}, \ref{tab:comparisonReentrancy}, and \ref{tab:comparisonTD}, respectively. The best results are highlighted in bold and the second best is marked with an underscore.

\begin{table}[htbp]
	\centering  % 显示位置为中间
	\caption{Performance of different models in terms of accuracy, recall, precision and F1-score on arithmetic vulnerability locating tasks.}  % 表格标题
	\label{tab:comparisonAri}  % 用于索引表格的标签
	\resizebox{\linewidth}{!}{
		\begin{tabular}{ccccc}
			\toprule
			& Acc(\%) & Recall(\%) & Precision(\%) & F1-score(\%) \\
			\midrule
			Transformer & 72.38 & 61.79 & 68.60 & 64.90 \\
			DGCNN & \underline{80.77} & 75.94 & 83.73 & 79.36 \\
			Peculiar & 65.20 & \underline{84.45} & 60.35 & 72.72 \\
			TMP & 80.00 & 81.00 & \underline{85.87} & \underline{83.36} \\
			MRN-GCN & \textbf{88.96} & \textbf{84.50} & \textbf{92.99} & \textbf{88.51} \\
			\bottomrule
		\end{tabular}
	}
\end{table}

\begin{table}[htbp]
	\centering  % 显示位置为中间
	\caption{Performance of different models on reentrancy vulnerability locating tasks.}  % 表格标题
	\label{tab:comparisonReentrancy}  % 用于索引表格的标签
	\resizebox{\linewidth}{!}{
		\begin{tabular}{ccccc}
			\toprule
			& Acc(\%) & Recall(\%) & Precision(\%) & F1-score(\%) \\
			\midrule
			Transformer & 89.63 & 86.99 & 85.68 & 86.29 \\
			DGCNN & 90.20 & \underline{89.49} & \underline{92.90} & \underline{90.92} \\
			Peculiar & \underline{92.36} & 89.29 & 91.19 & 90.42 \\
			TMP & 92.12 & 88.36 & 90.12 & 89.21 \\
			MRN-GCN & \textbf{96.43} & \textbf{98.18} & \textbf{97.07} & \textbf{97.62} \\
			\bottomrule
		\end{tabular}
	}
\end{table}

\begin{table}[htbp]
	\centering  % 显示位置为中间
	\caption{Performance of different models on timestamp dependency vulnerability locating tasks.}  % 表格标题
	\label{tab:comparisonTD}  % 用于索引表格的标签
	\resizebox{\linewidth}{!}{
		\begin{tabular}{ccccc}
			\toprule
			& Acc(\%) & Recall(\%) & Precision(\%) & F1-score(\%) \\
			\midrule
			Transformer & 85.03 & 81.85 & 82.08 & 81.83 \\
			DGCNN & \underline{88.36} & \underline{90.18} & \underline{87.07} & \underline{88.46} \\
			Peculiar & 79.23 & 82.60 & 78.16 & 79.65 \\
			TMP & 88.22 & 84.13 & 83.24 & 83.68 \\
			MRN-GCN & \textbf{94.21} & \textbf{92.81} & \textbf{94.69} & \textbf{93.71} \\
			\bottomrule
		\end{tabular}
	}
\end{table}

We first compare MRN-GCN with Transformer. It can be observed that MRN-GCN significantly improves the performance of the vulnerable function locating tasks over Transformer. Specifically, MRN-GCN outperforms Transformer on the accuracy of the three vulnerable function locating tasks by 16.6\%, 6.8\% and 9.2\%, respectively. MRN-GCN also achieves higher precision, recall and F1-score than Transformer by 16.1\%, 14.0\% and 15.6\% on average, respectively. The function code is converted to a sequence with Transformer. The sequence structure describes some of the semantic information in the smart contract function, but loses the important structural and control flow information in the function. Therefore, Transformer is unable to capture the structural and the logical features in the code. The experimental results demonstrate that the performance of the graph based models is generally better than that of Transformer which processes sequences.

Next, we compare MRN-GCN with the other graph based models, including DGCNN, Peculiar and TMP. MRN-GCN significantly improves the performance of the vulnerable function locating tasks compared to the three graph based baseline approaches. 

DGCNN learns the semantic and structural information of the code from the control flow graph (CFG). Note that the CFG lacks the representation of data flows. In addition, DGCNN only extracts the node features, while ignoring the edge features. That is, DGCNN neglects the features of different kinds of edges, such as data types, control information, etc. In contrast, our MRNG can characterize the complex relationships between data and instructions in the smart contract, and our MRN-GCN learns the MRNG using the edge-enhanced graph convolution network and self-attention mechanism. Consequently, our method can learn richer semantic and structural information than DGCNN. Tables \ref{tab:comparisonAri}$\sim$\ref{tab:comparisonTD} demonstrate that MRN-GCN obtains better results than DGCNN. Specifically, MRN-GCN outperforms DGCNN in terms of accuracy on the three vulnerable function locating tasks by 8.2\%, 6.2\% and 5.9\%, respectively. Furthermore, MRN-GCN achieves better precision, recall and F1-score performance than DGCNN by 7.0\%, 6.6\% and 9.2\% on average, respectively.

Peculiar builds a feature graph based on the crucial data flows in the smart contract and uses GNNs to extract features from the feature graph. However, the feature graph loses the control structures and the order of statement execution, etc. As a result, the detection performance varies significantly when using Peculiar to detect different vulnerabilities. Moreover, compared to CodeGraphBERT \cite{RN34} used by Peculiar, MRN-GCN uses an edge-enhanced graph convolution network, which allows MRN-GCN to better exploit the syntactic and semantic features inside functions in the smart contract. In addition, MRN-GCN employs a self-attention mechanism to balance local node features with global graph features. Therefore, MRN-GCN can effectively learn the relationships between data and instructions and hence improve the vulnerable function locating performance. Compared to Peculiar, MRN-GCN achieves better and more consistent results than Peculiar. Specifically, MRN-GCN improves the accuracy of the three vulnerable function locating tasks by 23.8\%, 4.1\% and 15.0\%, respectively. In addition, MRN-GCN increases the precision, recall and F1-score over Peculiar by 18.4\%, 6.4\% and 12.4\% on average, respectively.

TMP first constructs a contract graph based on the data and control dependencies in the code. TMP then normalizes the contract graph to highlight the major nodes and learns the features from the normalized graph for vulnerability detection. Note that some syntax information, such as variable types, variable values, etc., may be lost during the normalization. On the contrary, our MRNG contains more syntactic and semantic information than the normalized graph used by TMP. Our method creates the MRNG for a smart contract following 3 steps, i.e., building an AST for each function in a smart contract, generating the MRFG based on the AST by adding edges between nodes to describe different semantic and structural information, creating the MRNG with each node being an MRFG and each edge representing the invocation relationship between two corresponding functions. Furthermore, our MRN-GCN uses the edge-enhanced graph convolution network and self-attention mechanism to better extract the features from the MRNG. Accordingly, MRN-GCN achieves better performance than TMP. For instance, MRN-GCN obtains better accuracy than TMP by 8.5\%, 7.3\% and 6.0\%, respectively. MRN-GCN also outperforms TMP in terms of precision, recall and F1-score by 7.0\%, 6.6\% and 7.9\% on average for the three vulnerable function locating tasks, respectively.

In summary, the experimental results shown in Tables \ref{tab:comparisonAri}$\sim$\ref{tab:comparisonTD} demonstrate that the three graph based models, i.e., DGCNN, Peculiar and TMP, have the following two main issues. First, the graphs constructed by these three methods are unable to comprehensively distinguish different relationships between nodes, including the execution order, the data types of variables, control flows, data flows, etc. As a result, the learned features from the graphs are incomplete. Second, these three models lose or ignore some edge information when learning the graphs, such that these models cannot effectively extract the syntactic and semantic features from the graphs. In contrast, our MRNG retains the full data and control flows of each function, providing a fine-grained representation of the semantic and syntactic information inside the function. At the same time, our MRNG constructed based on inter-function invocation relationships represents the logical structure of the smart contract more accurately than the existing graph based methods. Therefore, the MRNG is more effective than the graphs used by the baseline methods. In addition, our MRN-GCN uses the edge-enhanced graph convolution network and self-attention mechanism to effectively capture the syntactic and semantic features, when learning the graph features. Accordingly, the proposed model MRN-GCN achieves the best performance in terms of accuracy, precision, recall and F1-score on all the three vulnerable function locating tasks.

The experimental results show that the proposed model MRN-GCN is stable as well as effective. Peculiar achieves the second best accuracy of 92.36\% on the reentrancy vulnerability locating task, but performs poorly on the arithmetic and timestamp dependency vulnerability locating tasks. TMP obtains the second highest precision and F1-score on the arithmetic vulnerability locating task. DGCNN ranks the second best on the timestamp dependency vulnerability locating task. DGCNN also obtains the second best accuracy on arithmetic vulnerability locating task and the second best recall, precision and F1-score on the reentrancy vulnerability locating task. DGCNN processes the MRNG for vulnerability locating, whereas the edges in each MRFG corresponding to a function only indicate whether the nodes have relationships. The experimental results demonstrate that the proposed MRNG can improve the vulnerability locating performance over the other graphs even without edge types in each MRFG. The MRNG can characterize the relationships between functions, in addition to the statement execution order, the control flows and the data flows in each function. Therefore, MRN-GCN can achieve superior performance.

%In general, the machine learning model under the classification task outputs the prediction probability, so when we use performance indicators such as accuracy rate and recall rate to evaluate the effectiveness of a model, we need to set a threshold to obtain the classification result through comparison with the threshold, such as setting the prediction probability greater than the threshold as a positive example, and setting the less than the threshold as a negativate case. The setting of thresholds affects the generalization ability of the model, and it also affects our judgment of the effectiveness of the model.

\begin{figure}[hbt]
	\centering
	\includegraphics[width=3.5in]{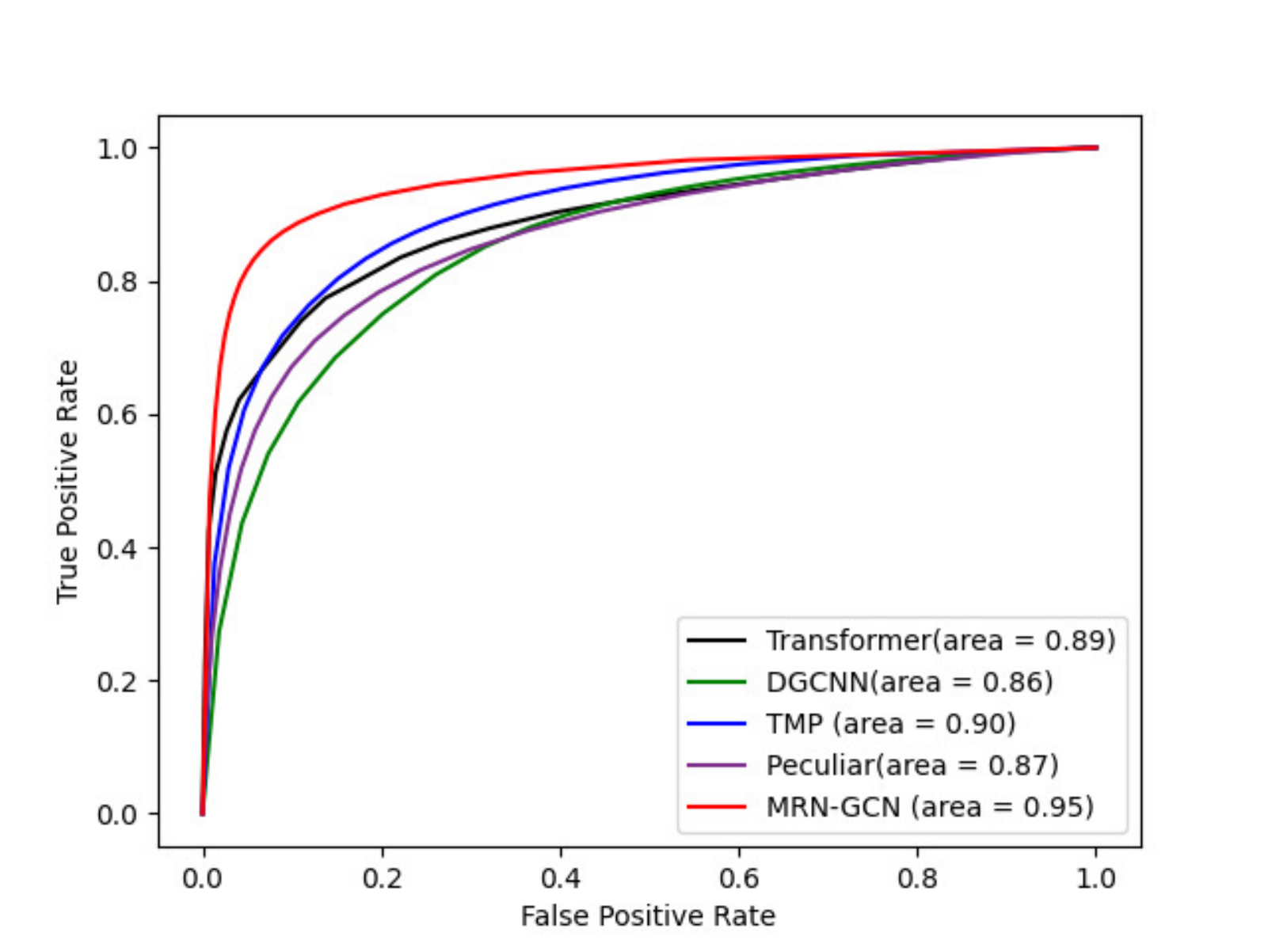}
	\caption{ROC curves of different models on arithmetic vulnerability locating tasks.}
	\label{fig:ari_roc}
\end{figure}

\begin{figure}[hbt]
	\centering
	\includegraphics[width=3.5in]{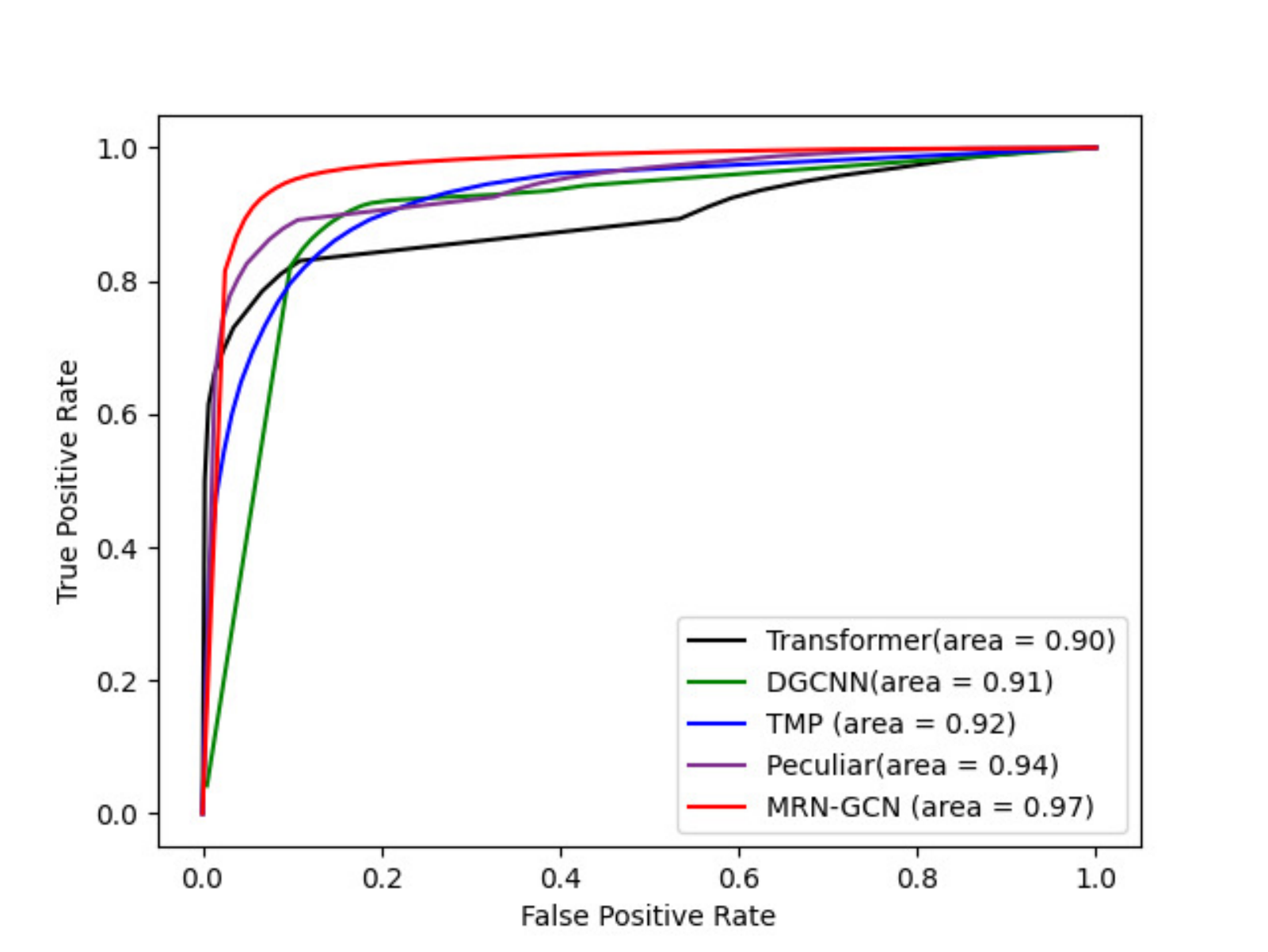}
	\caption{ROC curves of different models on reentrancy vulnerability locating tasks.}
	\label{fig:reentrancy_roc}
\end{figure}

\begin{figure}[hbt]
	\centering
	\includegraphics[width=3.5in]{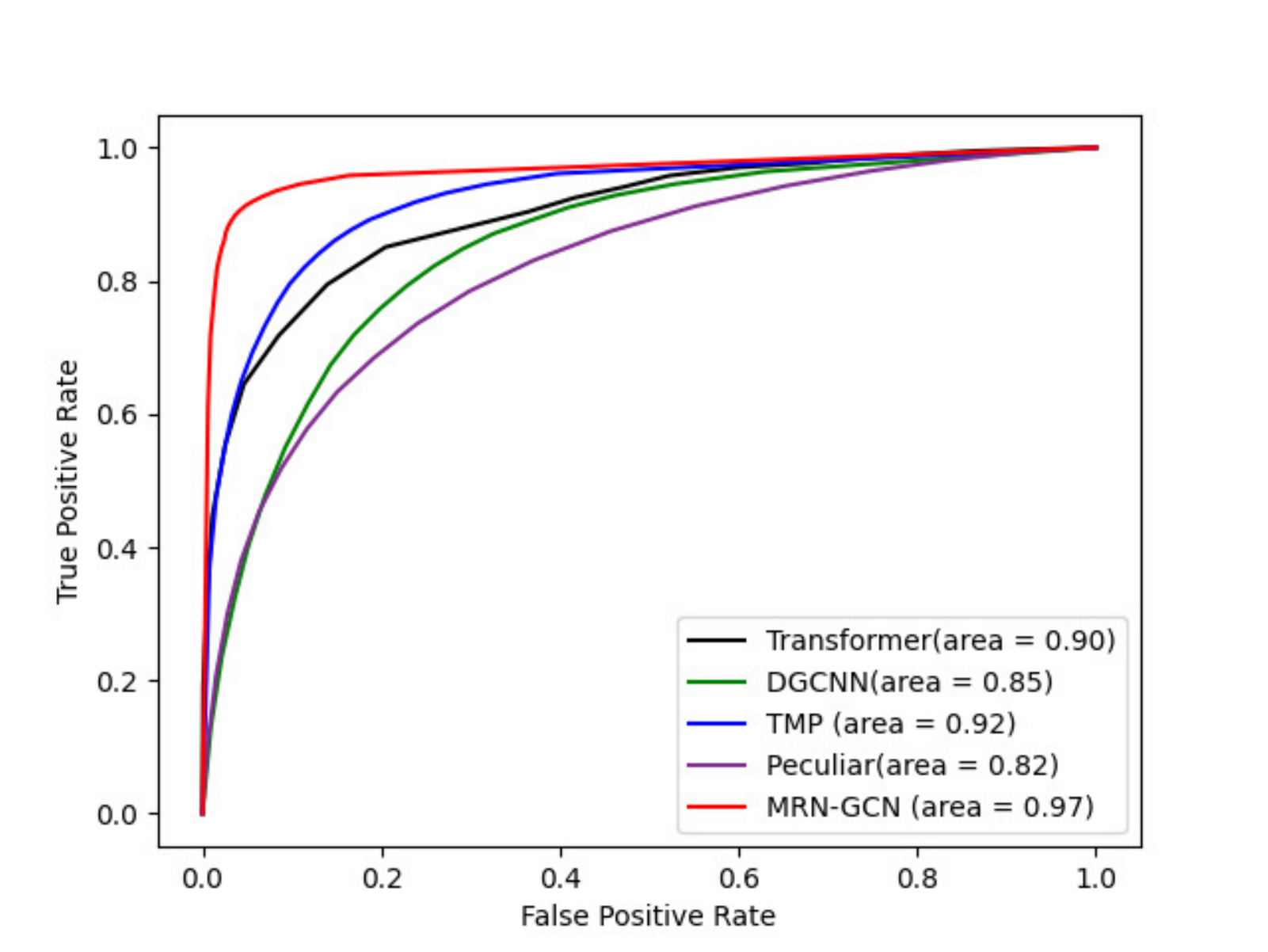}
	\caption{ROC curves of different models on timestamp vulnerability locating tasks.}
	\label{fig:TD_roc}
\end{figure}

We introduce receiver operating characteristics (ROC) curves to further evaluate the effectiveness of each model. ROC is a graph of true positive rate (TPR) and false positive rate (FPR) under different thresholds. The calculation of TPR and FPR is not affected by unbalanced data. Therefore,  ROC can still make a reasonable evaluation of the model performance even with unbalanced positive and negative samples. To quantitatively measure the ROC curve and evaluate the classifier's performance, the area under the ROC curve (AUC) is often used to measure the classifier's ability to distinguish between positive and negative samples. The ROC curves and AUC values of the different models on the arithmetic, reentrancy and timestamp dependecny vulnerability locating tasks are shown in Figs. \ref{fig:ari_roc}, \ref{fig:reentrancy_roc} and \ref{fig:TD_roc}, respectively. The experimental results show that MRN-GCN achieves the AUC of 0.95, 0.97 and 0.97 on the three vulnerability locating tasks, respectively, which are all the best among the different models.

%\begin{figure*}[htbp]
%	\centering  %图片全局居中
%	\subfloat[]{\includegraphics[width=0.34\textwidth]{SOURCE/fig/ari_roc.png}}
%	\subfloat[]{\includegraphics[width=0.34\textwidth]{SOURCE/fig/reen_roc.png}}
%	\subfloat[]{\includegraphics[width=0.34\textwidth]{SOURCE/fig/time_roc.png}}
%	\caption{(a). ROC curves of different methods on arithmetic vulnerability detection
	%		task. (b). ROC curves of different methods on reentrancy vulnerability detection
	%		task. (c). ROC curves of different methods on timestamp vulnerability detection
	%		task. }
%	\label{Fig.main}
%\end{figure*}
\subsubsection{Effect of the Multi-Relational Function Graph}
To investigate the effect of the proposed Multi-Relational Graph (MRFG), we compare the proposed model MRN-GCN with N-GCN, where N-GCN uses 0 or 1 to represent whether two nodes are connected within each function. That is, the edge types are not described in N-GCN. Correspondingly, we extract N-GCN features using Graph Attention Network (GAT) \cite{RN33} .

As illustrated in Tables \ref{tab:ari_gat}$\sim$\ref{tab:time_gat}, model MRN-GCN outperforms model N-GCN in terms of accuracy, precision, recall and F1-score on all the three vulnerable function locating tasks. Specifically, MRN-GCN improves the accuracy on the arithmetic, reentrancy and timestamp dependency vulnerability locating tasks by 7.8\%, 3.2\% and 10.4\% over N-GCN, respectively. MRN-GCN also achieves better performance of precision, recall and F1-score than N-GCN on the three vulnerable function locating tasks by an average of 8.4\%, 5.8\% and 6.7\%, respectively. Each edge in the N-GCN only indicates that two nodes have some relationship, but fails to differentiate the relationships between node pairs. In contrast, the proposed MRFG characterizes the structure, the execution order, the control and the data flows within each function. Different types of edges in the MRFG indicate different relationships between nodes, including the order of statement execution, the sequence of operators, variable types, value assignment, fallback, etc. Consequently, the proposed MRFG contains more structural and semantic information in the code, which enables neural networks to understand the smart contract code effectively and improves the performance of vulnerable function locating tasks. 

\begin{table}[htbp]
	\centering  % 显示位置为中间
	\caption{Performance of different models on arithmetic vulnerability locating tasks.}  % 表格标题
	\label{tab:ari_gat}  % 用于索引表格的标签
	\resizebox{\linewidth}{!}{
		\begin{tabular}{ccccc}
			\toprule
			& Acc(\%) & Recall(\%) & Precision(\%) & F1-score(\%) \\
			\midrule
			N-GCN & 81.20 & 79.59 & 83.12 & 80.97 \\
			MRN-GCN & 88.96 & 84.50 & 92.99 & 88.51 \\
			\bottomrule
		\end{tabular}
	}
\end{table}

\begin{table}[htbp]
	\centering  % 显示位置为中间
	\caption{Performance of different models on reentrancy vulnerability locating tasks.}  % 表格标题
	\label{tab:reen_gat}  % 用于索引表格的标签
	\resizebox{\linewidth}{!}{
		\begin{tabular}{ccccc}
			\toprule
			& Acc(\%) & Recall(\%) & Precision(\%) & F1-score(\%) \\
			\midrule
			N-GCN & 93.19 & 93.17 & 94.61 & 95.34 \\
			MRN-GCN & 96.43 & 98.18 & 97.07 & 97.62 \\
			\bottomrule
		\end{tabular}
	}
\end{table}

\begin{table}[htbp]
	\centering  % 显示位置为中间
	\caption{Performance of different models on timestamp dependency vulnerability locating tasks.}  % 表格标题
	\label{tab:time_gat}  % 用于索引表格的标签
	\resizebox{\linewidth}{!}{
		\begin{tabular}{ccccc}
			\toprule
			& Acc(\%) & Recall(\%) & Precision(\%) & F1-score(\%) \\
			\midrule
			N-GCN & 83.83 & 85.31 & 81.93 & 83.34 \\
			MRN-GCN & 94.21 & 92.81 & 94.69 & 93.71 \\
			\bottomrule
		\end{tabular}
	}
\end{table}

\subsubsection{Effect of Different Modules in MRN-GCN}
The feature extraction in our proposed model MRN-GCN consists of three important parts: the edge-enhanced graph convolutional module, the self-attention module and the nested graph convolutional module. The edge-enhanced graph convolutional module performs feature extraction from MRFGs. The module enables the semantic information between neighboring nodes and the control flow information between the nodes and edges in MRFGs to be fully propagated, which enhances the extraction of function features. The self-attention mechanism further extracts and aggregates the output features obtained by each layer in the edge-enhanced graph convolutional module. The nested graph convolutional module captures the association between functions according to the invocation relationships between functions in the smart contract to promote the extraction of code features.

In order to investigate the impacts of the two components of self-attention and nested graph convolution in our proposed model, we compare MRN-GCN with MRN-GCN (w/o SA), MR-GCN and MR-GCN (w/o SA). Among them, MRN-GCN (w/o SA) removes the self-attention module, MR-GCN ignores the nested graph convolutional module, and MR-GCN (w/o SA) excludes both the self-attention module and the nested graph convolutional module. Tables \ref{tab:ablationArithmetic}, \ref{tab:ablationReentrancy} and \ref{tab:ablationTD} show the experimental results of arithmetic, reentrancy and timestamp dependency vulnerability locating tasks, respectively. It can be observed that model MRN-GCN achieves the best performance in terms of accuracy, precision, recall and F1-score.

\begin{table}[htbp]
	\centering  % 显示位置为中间
	\caption{Performance of different models on arithmetic vulnerability locating tasks.}  % 表格标题
	\label{tab:ablationArithmetic}  % 用于索引表格的标签
	\resizebox{\linewidth}{!}{
	\begin{tabular}{ccccc}
		\toprule
		& Acc(\%) & Recall(\%) & Precision(\%) & F1-score(\%) \\
		\midrule
		MR-GCN(w/o   SA) & 80.62 & 81.02 & 80.80 & 81.84 \\
		MR-GCN & 83.77 & 82.39 & 86.24 & 84.21 \\
		MRN-GCN(w/o   SA) & 82.35 & 82.30 & 82.87 & 82.45 \\
		MRN-GCN & 88.96 & 84.50 & 92.99 & 88.51 \\
		\bottomrule
	\end{tabular}
	}
\end{table}
\begin{table}[htbp]
	\centering  % 显示位置为中间
	\caption{Performance of different models on reentrancy vulnerability locating tasks.}  % 表格标题
	\label{tab:ablationReentrancy}  % 用于索引表格的标签
	\resizebox{\linewidth}{!}{
	\begin{tabular}{ccccc}
		\toprule
		& Acc(\%) & Recall(\%) & Precision(\%) & F1-score(\%) \\
		\midrule
		MR-GCN (w/o SA) & 87.70 & 89.92 & 85.06 & 87.40 \\
		MR-GCN & 91.79 & 90.93 & 91.78 & 91.34 \\
		MRN-GCN (w/o SA) & 93.43 & 95.40 & 95.86 & 95.60 \\
		MRN-GCN & 96.43 & 98.18 & 97.07 & 97.62 \\
		\bottomrule
	\end{tabular}
	}
\end{table}
\begin{table}[htbp]
	\centering  % 显示位置为中间
	\caption{Performance of different models on timestamp dependency vulnerability locating tasks.}  % 表格标题
	\label{tab:ablationTD}  % 用于索引表格的标签
	\resizebox{\linewidth}{!}{
	\begin{tabular}{ccccc}
		\toprule
		& Acc(\%) & Recall(\%) & Precision(\%) & F1-score(\%) \\
		\midrule
		MR-GCN (w/o SA) & 85.53 & 87.26 & 82.77 & 84.94 \\
		MR-GCN & 89.46 & 89.84 & 87.95 & 88.87 \\
		MRN-GCN (w/o SA) & 88.72 & 88.22 & 91.20 & 87.50 \\
		MRN-GCN & 94.21 & 92.81 & 94.69 & 93.71 \\
		\bottomrule
	\end{tabular}
	}
\end{table}

The self-attention module fuses the features obtained by each layer in the graph convolutional module, such that the finally obtained features can achieve a balance between each node itself and the relevant nodes with different distances. When the nested contract graph is included in the model for the three vulnerable function locating tasks, MRN-GCN respectively improves the accuracy by 6.6\%, 3.0\% and 5.5\% over MRN-GCN (w/o SA), and MRN-GCN outperforms MRN-GCN (w/o SA) on precision, recall and F1-score by the average of 4.94\%, 3.19\% and 4.76\%, respectively. When the nested contract graph is missing in the model for the three vulnerable function locating tasks, MR-GCN respectively achieves higher accuracy than MR-GCN (w/o SA) by 3.2\%, 4.1\% and 3.9\%, and MR-GCN obtains better precision, recall and F1-score than MR-GCN (w/o SA) by the average of 5.8\%, 1.6\% and 3.4\%, respectively. The experimental results demonstrate that the self-attention module can improve the performance no matter whether the nested contract graph is utilized in the model or not.

The nested contract graph module can describe the relationships between smart contract functions. Each node in the nested contract graph is an MRFG, and each edge describes the relationship between two functions, i.e., function invocations. The module learns all the functions jointly, such that the vulnerabilities caused by function invocations can be detected by the neural network. When the self-attention mechanism is included in the model for the three vulnerable function locating tasks, MRN-GCN respectively improves the accuracy over MR-GCN by 5.2\%, 4.6\% and 4.8\%, and MRN-GCN also outperforms MR-GCN in terms of precision, recall and F1-score by the average of 6.3\%, 4.1\% and 5.1\%, respectively.  When the self-attention mechanism is excluded from the model for the three vulnerable function locating tasks, MRN-GCN (w/o SA) achieves higher accuracy than MR-GCN (w/o SA) by 1.7\%, 5.7\% and 3.2\%, respectively, and MRN-GCN (w/o SA) respectively obtains better precision, recall and F1-score than MR-GCN (w/o SA) by the average of 6.9\%, 2.6\% and 3.6\%. The experimental results illustrate that the nested contract graph module can improve the performance no matter whether the self-attention mechanism is incorporated in the model or not.

In general, both the self-attention module and the nested graph convolutional module are of great importance in our proposed model MRN-GCN. The self-attention module can adaptively extract the features within each function, and the nested graph convolutional module can effectively extract the latent features between smart contract functions. 

\section{Threats to Validity}
\label{sect:threats}
In this section, we discuss the limitations of our study in terms of both internal and external threats.

\subsection{External Threats}
The main external threat to our study is the reliability of the datasets. We use real-world smart contracts from SmartBugs \cite{RN27}. Most of the smart contracts in SmartBugs are not labelled. In the experiments, we manually label the datasets using a voting strategy. Specifically, we use multiple detection tools \cite{torres2018osiris,RN40,RN53,RN57,mythril} to check the possible vulnerabilities within smart contract functions. If the majority of the detection tools determine that a function has a specific kind of vulnerability, we mark the function's label as that kind of vulnerability. However, most smart contract detection tools are not absolutely accurate. Accordingly, there may be some labelling errors in the tagging process, which threatens the reliability of the datasets. In the future, we will consider the use of expert knowledge and semi-supervised learning methods to deal with the unreliable datasets.

\subsection{Internal Threats}
Our method faces internal threats as well. When training the model, the settings of hyperparameters and the randomness of model initialization parameters affect the model performance. We can use grid search to find the appropriate hyperparameters, and we may perform multiple rounds of experiments to reduce the impact of parameter randomness. Furthermore, in order to strike the balance between positive and negative samples in each training set, which affects the generalization ability of the model, we shuffle the dataset to mitigate this potential threat. In addition, the size of the dataset used can also affect the classification ability of the model and hence the effectiveness of the method. A small dataset is prone to overfitting, resulting in the poor performance of the model. In general, a large dataset can help the neural network to learn the features of smart contract code accurately and can hence reduce the risk of overfitting. Therefore, it is better to choose a large dataset with balanced positive and negative samples.

\section{Conclusion}
\label{sect:conclusion}
In this paper, we studied the problem of vulnerable smart contract function locating based on the smart contract source code. We proposed a novel Multi-Relational Nested contract Graph (MRNG) to characterize the rich syntax and semantic information in the smart contract code, including the order of statement execution, control flows and data flows. The nodes and edges in the MRNG represent functions and call relationships between functions, respectively. Each node in MRNG contains a Multi-Relational Function Graph (MRFG) which represents the corresponding function code. Each MRFG uses different types of edges to represent the different control and data relationships between nodes within the function. We also proposed a Multi-Relational Nested Graph Convolutional Network (MRN-GCN) to process the MRNG. MRN-GCN first extracts and aggregates features from the MRFG corresponding to each function in the smart contract, using the edge-enhanced graph convolution network and self-attention mechanism. The extracted feature vector is then assigned to each node in the nested contract graph MRNG to obtain a new Featured Contract Graph (FCG). The graph convolution network is used to further extract features from the FCG. Finally, a feed forward network with a Sigmoid function is used to locate the vulnerable functions. The experimental results on the real-world smart contract datasets showed that model MRN-GCN could effectively improve the accuracy, precision, recall and F1-score performance of vulnerable smart contract function locating. Specifically, MRN-GCN achieves 88.96\%, 96.43\% and 94.21\% accuracy on arithmetic, reentrancy and timestamp dependency vulnerability locating tasks, respectively, and obtains 94.92\%, 91.83\% and 94.21\% on average in terms of precision, recall and F1-score, respectively. The proposed MRNG enables neural networks to understand the smart contract code effectively and improves the performance of vulnerable function locating tasks. Furthermore, both the self-attention module and the nested graph convolutional module are of great importance in our proposed model MRN-GCN.
\bibliographystyle{elsarticle-num} %声明选择的格式
\bibliography{references} %bib文件名，需要放在同一个文件夹下，否则要在filename前说明路径

\end{document}